\RequirePackage{fix-cm}
\documentclass[twocolumn,epjc3]{svjour3}  
\usepackage{newtxtext,newtxmath} 
\RequirePackage{graphicx}
\RequirePackage{subfigure}
\RequirePackage{rotating}
\RequirePackage{pdflscape}
\RequirePackage{color}
\usepackage{mathtools}
\usepackage{extarrows}
\usepackage{bm}
\RequirePackage{latexsym}
\RequirePackage{xspace}
\RequirePackage[numbers,sort&compress]{natbib}
\RequirePackage[colorlinks,citecolor=blue,urlcolor=blue,linkcolor=blue]{hyperref}
\usepackage{amsmath,amsfonts,bm}
\usepackage{wasysym}
\usepackage{color}
\usepackage{lineno}
\usepackage[usenames,dvipsnames,svgnames,table]{xcolor}	
\usepackage[nottoc, notlof, notlot]{tocbibind} 
\let\oldequation\equation
\let\oldendequation\endequation

\renewenvironment{equation}
  {\linenomathNonumbers\oldequation}
  {\oldendequation\endlinenomath}

\let\oldalign\align
\let\oldendalign\endalign

\renewenvironment{align}
  {\linenomathNonumbers\oldalign}
  {\oldendalign\endlinenomath}

\usepackage{acronym}
\usepackage{xcolor}

\newcommand{\M}{$\mathcal{M}$}
\newcommand{\onbb}{$0\nu\beta\beta$\xspace}
\newcommand{\nnbb}{$2\nu\beta\beta$\xspace}

\newcommand{\natlmo}{Li$_2$MoO$_4$\xspace}
\newcommand{\LMO}{Li$_{2}${}$^{100}$MoO$_4$\xspace}
\newcommand{\edw}{EDELWEISS\xspace}

\journalname{Eur. Phys. J. C}

\begin{document}

\title{The background model of the CUPID-Mo \onbb experiment}

\author{
C.~Augier\thanksref{IPNL}\and
A.~S.~Barabash\thanksref{ITEP}\and
F.~Bellini\thanksref{Sapienza,INFN-Roma}\and
G.~Benato\thanksref{GSSI, LNGS}\and
M.~Beretta\thanksref{UCB}\and
L.~Berg\'e\thanksref{IJCLab}\and
J.~Billard\thanksref{IPNL}\and
Yu.~A.~Borovlev\thanksref{NIIC}\and
L.~Cardani\thanksref{INFN-Roma}\and
N.~Casali\thanksref{INFN-Roma}\and
A.~Cazes\thanksref{IPNL}\and
E.~Celi\thanksref{GSSI, LNGS}\and
M.~Chapellier\thanksref{IJCLab}\and
D.~Chiesa\thanksref{Milano,INFN-Milano}\and
I.~Dafinei\thanksref{INFN-Roma}\and
F.~A.~Danevich\thanksref{KINR,INFN-Roma}\and
M.~De~Jesus\thanksref{IPNL}\and
P.~de~Marcillac\thanksref{IJCLab}\and
T.~Dixon\thanksref{IJCLab}\and
L.~Dumoulin\thanksref{IJCLab}\and
K.~Eitel\thanksref{KIT-IK}\and
F.~Ferri\thanksref{CEA-IRFU}\and
B.~K.~Fujikawa\thanksref{LBNLNSD}\and
J.~Gascon\thanksref{IPNL}\and
L.~Gironi\thanksref{Milano,INFN-Milano}\and
A.~Giuliani\thanksref{IJCLab}\and
V.~D.~Grigorieva\thanksref{NIIC}\and
M.~Gros\thanksref{CEA-IRFU}\and
D.~L.~Helis\thanksref{GSSI, LNGS}\and
H.~Z.~Huang\thanksref{Fudan}\and
R.~Huang\thanksref{UCB}\and
L.~Imbert\thanksref{e1,IJCLab}\and
J.~Johnston\thanksref{MIT}\and
A.~Juillard\thanksref{IPNL}\and
H.~Khalife\thanksref{CEA-IRFU}\and 
M.~Kleifges\thanksref{KIT-IPE}\and
V.~V.~Kobychev\thanksref{KINR}\and
Yu.~G.~Kolomensky\thanksref{UCB,LBNLNSD}\and
S.I.~Konovalov\thanksref{ITEP}\and
J. ~Kotila\thanksref{Jyv,Jyv2,yale}\and
P.~Loaiza\thanksref{IJCLab}\and
L.~Ma\thanksref{Fudan}\and
E.~P.~Makarov\thanksref{NIIC}\and
R.~Mariam\thanksref{IJCLab}\and
L.~Marini\thanksref{UCB,GSSI}\and
S.~Marnieros\thanksref{IJCLab}\and
X.-F.~Navick\thanksref{CEA-IRFU}\and
C.~Nones\thanksref{CEA-IRFU}\and
E.B.~Norman\thanksref{UCBNE}\and
E.~Olivieri\thanksref{IJCLab}\and
J.~L.~Ouellet\thanksref{MIT}\and
L.~Pagnanini\thanksref{GSSI,LNGS}\and
L.~Pattavina\thanksref{LNGS,TUM}\and
B.~Paul\thanksref{CEA-IRFU}\and
M.~Pavan\thanksref{Milano,INFN-Milano}\and
H.~Peng\thanksref{USTC}\and
G.~Pessina\thanksref{INFN-Milano}\and
S.~Pirro\thanksref{LNGS}\and
D.~V.~Poda\thanksref{IJCLab}\and
O.~G.~Polischuk\thanksref{KINR,INFN-Roma}\and
S.~Pozzi\thanksref{INFN-Milano}\and
E.~Previtali\thanksref{Milano,INFN-Milano}\and
Th.~Redon\thanksref{IJCLab}\and
A.~Rojas\thanksref{LSM}\and
S.~Rozov\thanksref{JINR}\and 
V.~Sanglard\thanksref{IPNL}\and
J.A.~Scarpaci\thanksref{IJCLab}\and
B.~Schmidt\thanksref{CEA-IRFU}\and
Y.~Shen\thanksref{Fudan}\and
V.~N.~Shlegel\thanksref{NIIC}\and
V.~Singh\thanksref{UCB}\and
C.~Tomei\thanksref{INFN-Roma}\and
V.~I.~Tretyak\thanksref{KINR, LNGS}\and 
V.~I.~Umatov\thanksref{ITEP}\and
L.~Vagneron\thanksref{IPNL}\and
M.~Vel\'azquez\thanksref{UGA}\and
B.~Welliver\thanksref{UCB}\and
L.~Winslow\thanksref{MIT}\and
M.~Xue\thanksref{USTC}\and
E.~Yakushev\thanksref{JINR}\and
M.~Zarytskyy\thanksref{KINR}\and
A.~S.~Zolotarova\thanksref{CEA-IRFU}
}

\thankstext{e1}{e-mail: leonard.imbert@ijclab.in2p3.fr}

\institute{
Universit\'{e} Lyon 1, CNRS/IN2P3, IP2I-Lyon, F-69622, Villeurbanne, France  \label{IPNL} \and
National Research Centre Kurchatov Institute, Kurchatov Complex of Theoretical and Experimental Physics, 117218 Moscow, Russia \label{ITEP} \and 
Dipartimento di Fisica, Sapienza Universit\`a di Roma, P.le Aldo Moro 2, I-00185, Rome, Italy \label{Sapienza} \and 
INFN, Sezione di Roma Tor Vergata, Via della Ricerca Scientifica 1, I-00133, Rome, Italy \label{INFN-Roma} \and
Gran Sasso Science Institute, L'Aquila I-67100, Italy \label{GSSI} \and 
INFN, Laboratori Nazionali del Gran Sasso, I-67100 Assergi (AQ), Italy \label{LNGS} \and
Department of Physics, University of California, Berkeley, California 94720, USA \label{UCB} \and
Universit\'{e} Paris-Saclay, CNRS/IN2P3, IJCLab, 91405 Orsay, France \label{IJCLab} \and
Nikolaev Institute of Inorganic Chemistry, 630090 Novosibirsk, Russia \label{NIIC} \and
Dipartimento di Fisica, Universit\`{a} di Milano-Bicocca, I-20126 Milano, Italy \label{Milano} \and 
INFN, Sezione di Milano-Bicocca, I-20126 Milano, Italy \label{INFN-Milano} \and 
Institute for Nuclear Research of NASU, 03028 Kyiv, Ukraine \label{KINR} \and 
Institute for Astroparticle Physics, Karlsruhe Institute of Technology, 76021 Karlsruhe, Germany \label{KIT-IK} \and
Nuclear Science Division, Lawrence Berkeley National Laboratory, Berkeley, California 94720, USA \label{LBNLNSD} \and
IRFU, CEA, Universit\'{e} Paris-Saclay, F-91191 Gif-sur-Yvette, France  \label{CEA-IRFU} \and
Key Laboratory of Nuclear Physics and Ion-beam Application (MOE), Fudan University, Shanghai 200433, PR China \label{Fudan} \and
Massachusetts Institute of Technology, Cambridge, MA 02139, USA \label{MIT} \and
Institute for Data Processing and Electronics, Karlsruhe Institute of Technology, 76021 Karlsruhe, Germany \label{KIT-IPE} \and
 Department of Physics, University of Jyv\"askyl\"a, PO Box 35, FI-40014, Jyv\"askyl\"a, Finland\label{Jyv} 
\and Finnish Institute for Educational Research, University of Jyv\"askyl\"a, P.O. Box 35, FI-40014 Jyv\"askyl\"a, Finland \label{Jyv2} 
\and Center for Theoretical Physics, Sloane Physics Laboratory, Yale University, New Haven, Connecticut 06520-8120, USA \label{yale}\and
Department of Nuclear Engineering, University of California, Berkeley, California 94720, USA \label{UCBNE} \and
Physik Department, Technische Universit\"at M\"unchen, Garching D-85748, Germany \label{TUM} \and
Department of Modern Physics, University of Science and Technology of China, Hefei 230027, PR China \label{USTC} \and
LSM, Laboratoire Souterrain de Modane, 73500 Modane, France \label{LSM} \and
Laboratory of Nuclear Problems, JINR, 141980 Dubna, Moscow region, Russia \label{JINR} \and
Universit\'e Grenoble Alpes, CNRS, Grenoble INP, SIMAP, 38402 Saint Martin d'H\'eres, France \label{UGA}
}




\date{Received: date / Accepted: date}

\maketitle

\begin{abstract}
CUPID-Mo, located in the Laboratoire Souterrain de Modane (France), was a demonstrator for the next generation $0\nu\beta\beta$ decay experiment, CUPID. It consisted of an array of 20 enriched \LMO bolometers and 20 Ge light detectors and has demonstrated that the technology of scintillating bolometers with particle identification capabilities is mature. 
  Furthermore, CUPID-Mo can inform and validate the background prediction for CUPID. In this paper, we present a detailed model of the CUPID-Mo backgrounds. This model is able to describe well the features of the experimental data and enables studies of the \nnbb decay and other processes with high precision.  We also measure the radio-purity of the \LMO crystals which are found to be sufficient for the CUPID goals. Finally, we also obtain a background index in the region of interest of 3.7~$^{+0.9}_{-0.8}$~\text{(stat)}$^{+1.5}_{-0.7}$~\text{(syst)}~$\times~10 ^{-3}$~\text{counts/$\Delta E_{\text{FWHM}}$/mol$_{\text{iso}}$/yr}, the lowest in a bolometric $0\nu\beta\beta$ decay experiment.


\end{abstract}
\tableofcontents
\section{Introduction}
\label{intro} 
Neutrinoless double beta decay ($0\nu\beta\beta$)
is a hypothetical nuclear transition that would occur if the neutrino is its own antiparticle, or a  Majorana particle. It consists in the transformation of an even-even nucleus into a lighter isobar containing two more protons accompanied by the emission of two electrons and no other particles, with a change of the total lepton number by two units. 
The detection of this "matter-creating" process would represent  the observation of a new phenomenon beyond the Standard Model \cite{Agostini2022}.
Current best limits for $0\nu\beta\beta$ half-life are of the order of 10$^{24}$ -- 10$^{26}$ yr \cite{KLZ_IO,GERDA_final,CUORE,Azzolini:2022,Augier:2022,EXO,MJD}.
\\ \indent
The Standard Model process, two-neutrino double beta decay, \nnbb, includes also the emission of two $\bar{\nu}_e$ and  conserves lepton number. 
 Unlike $0\nu\beta\beta$ decay, \nnbb has a continuous energy spectrum and has been observed in more than ten nuclei with half-lives in the range of 10$^{18}$ -- 10$^{24}$ yr \cite{Barabash:2b2n}. 
 \\ \indent  One of the largest challenges in \onbb decay experiments is the control of the radioactive background, that may produce events in the signal energy region. These could mimic the very rare $0\nu\beta\beta$  signal 
 reducing the experimental sensitivity.
\\ \indent  During the last 10 years the scintillating bolometer technology has proved that bolometers based on lithium molibdate (\natlmo ), are very promising detectors for next generation \onbb searches \cite{LUMINEU,Armengaud:2021}. Scintillating bolometers were developed to reduce the background observed in the current leading \onbb bolometric experiment, CUORE \cite{CUORE_description}. In CUORE, the background in the region of interest 
is dominated by surface $\alpha$'s emitted from the copper structure holding the detectors \cite{CUORE_backg}. The array of 988 TeO$_2$ bolometers, installed at the Laboratori Nazionali del Gran Sasso, Italy, has observed a  background in the $^{130}$Te region of interest ($\mathcal Q_{\beta\beta}$ = 2527 keV) of $(1.49\pm 0.04 )\times 10^{-2}\ \mathrm{counts/keV/kg/yr}$ \cite{CUORE, CUORE:2021, Alduino:2017}. The next generation experiment CUPID (Cuore Upgrade with Particle IDentification) will drastically reduce the background thanks to the simultaneous readout of heat and light signals. The capability to discriminate  $\beta/\gamma$ from $\alpha$ particles with scintillating bolometers relies on the fact that the light emitted in the \LMO by  $\alpha$ particles is about a factor 5  smaller compared to the light emitted by $\beta/\gamma$'s of the same energy \cite{LUMINEU, Denys_review}.  
 \\ \indent In addition to the particle discrimination, the CUPID strategy to reduce the background relies on the radiopurity of the scintillating crystals and the minimisation of the passive materials \cite{CUPID_pre-CDR}. Another key point is that the $\mathcal Q_{\beta\beta}$ energy value of $^{100}$Mo (3034 keV) is higher than 2615~keV, implying a signal located above the majority of $\gamma$ lines from natural radioactivity. 
 \\ \indent
  The CUPID-Mo experiment \cite{Armengaud:2021}, located in the Laboratoire Souterrain de Modane (LSM) in France, under an overburden of 4800 metres water equivalent, was built as a demonstrator experiment for CUPID. It consisted of 20 \LMO (LMO) scintillating bolometers and 20 Ge light detectors (LDs) for a simultaneous read-out of heat and light. 
One of the aims of CUPID-Mo was to validate the background predictions for CUPID, in particular the LMO crystal radiopurity and residual $\alpha$ background. 
\\ \indent In this paper we present the background model which describes the background sources in the CUPID-Mo experiment. This model is based on fitting the CUPID-Mo data to detailed Monte Carlo simulations. 
We show that the residual $\alpha$ background contribution and the radiopurity of the LMO crystals are sufficient to meet the CUPID background goal.

CUPID-Mo was also an important experiment in its own right. In particular, it has set the world-leading limits on the half-life of $0\nu\beta\beta$ decay of $^{100}$Mo to both ground and excited states \cite{Armengaud:2021, Augier:2022,ES-paper}.
The detailed study of the experimental backgrounds in the CUPID-Mo experiment enables a high precision measurement of the \nnbb decay rate and allows to disentangle between the Single State Dominance (SSD) or High State Dominance (HSD) mechanisms \cite{SSD_HSD1,SSD_HSD2} of the \nnbb decay process in $^{100}$Mo. It  also provides the basis to study new physics processes outside the Standard Model, which could distort the spectral shape of the \nnbb spectrum, such as $0\nu\beta\beta$ decay with emisson of Majoron(s), $2\nu\beta\beta$ decay with emisson of Bosonic neutrinos, Lorentz invariance violation or sterile neutrinos \cite{exoticprocess1,exoticprocess2,exoticprocess3,exoticprocess4,exoticprocess5,exoticprocess6,exoticprocess7, bosonicn, CPviolation, sterilen1}.

\section{The CUPID-Mo experiment}
 CUPID-Mo was installed in the EDELWEISS cryogenic set-up \cite{Armengaud:2017} at LSM. The experiment was in operation between March 2019 and June 2020. 
 
 \begin{figure}[htbp]
   \centering
    \includegraphics[width=0.5\textwidth]{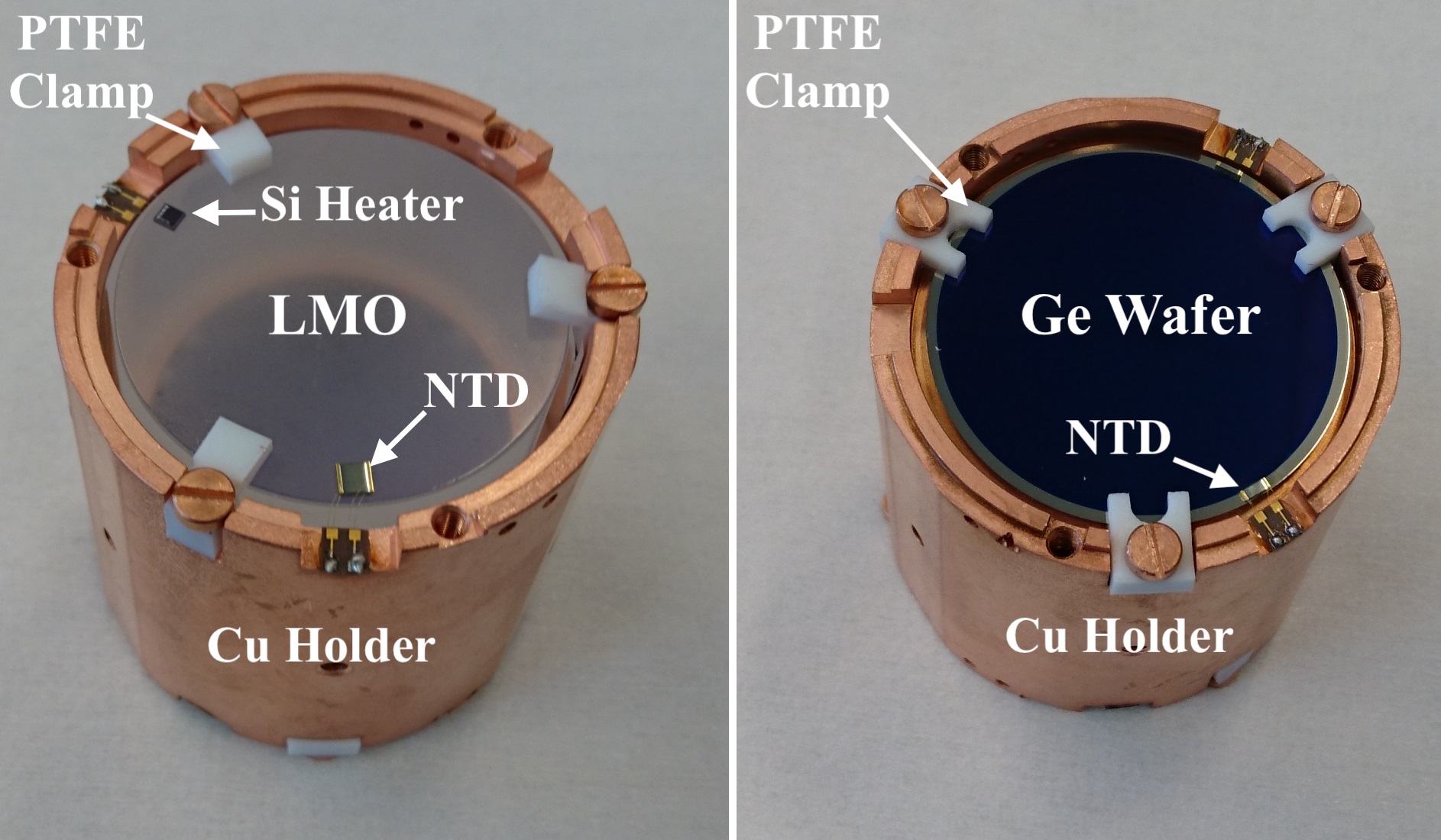}
    \caption{{\it Left}: An individual CUPID-Mo bolometer showing the transparent \LMO crystal, the copper holder, the NTD-Ge thermometer and the Teflon clamps. {\it Right}: View of the opposite side of the detector, showing the black light detector, fabricated from Ge wafers \cite{Augier:2022}.
    }
    \label{fig:CUPIDMo-pic}
\end{figure}
  \indent  CUPID-Mo used $^{100}$Mo-enriched LMO crystals, where the $^{100}$Mo, the double beta isotope, has been enriched at $\sim$~97\%. The basic detection modules are the crystals coupled to thermal sensors, consisting of a Neutron Transmutation Doped Ge thermistor, NTD. The top and bottom of the crystals are facing light detectors fabricated from Ge wafers, also instrumented with NTDs to readout the scintillation light signal.
The crystals are housed in cylindrical copper holders and supported by PTFE pieces, as shown in Fig.~\ref{fig:CUPIDMo-pic}. A reflective foil (3M Vikuiti\textregistered ) is installed around the crystals, inside the copper holder, to increase the light collection. The average weight of the CUPID-Mo crystals is 210 g and the total mass is 4.158 kg corresponding to 2.264 kg of 
$^{100}$Mo.  
\begin{figure*}
   \includegraphics[height=0.42\textwidth]{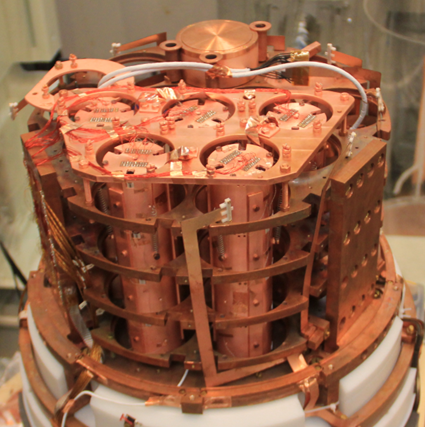}
   \hspace{2cm}
    \includegraphics[height=0.42\textwidth]{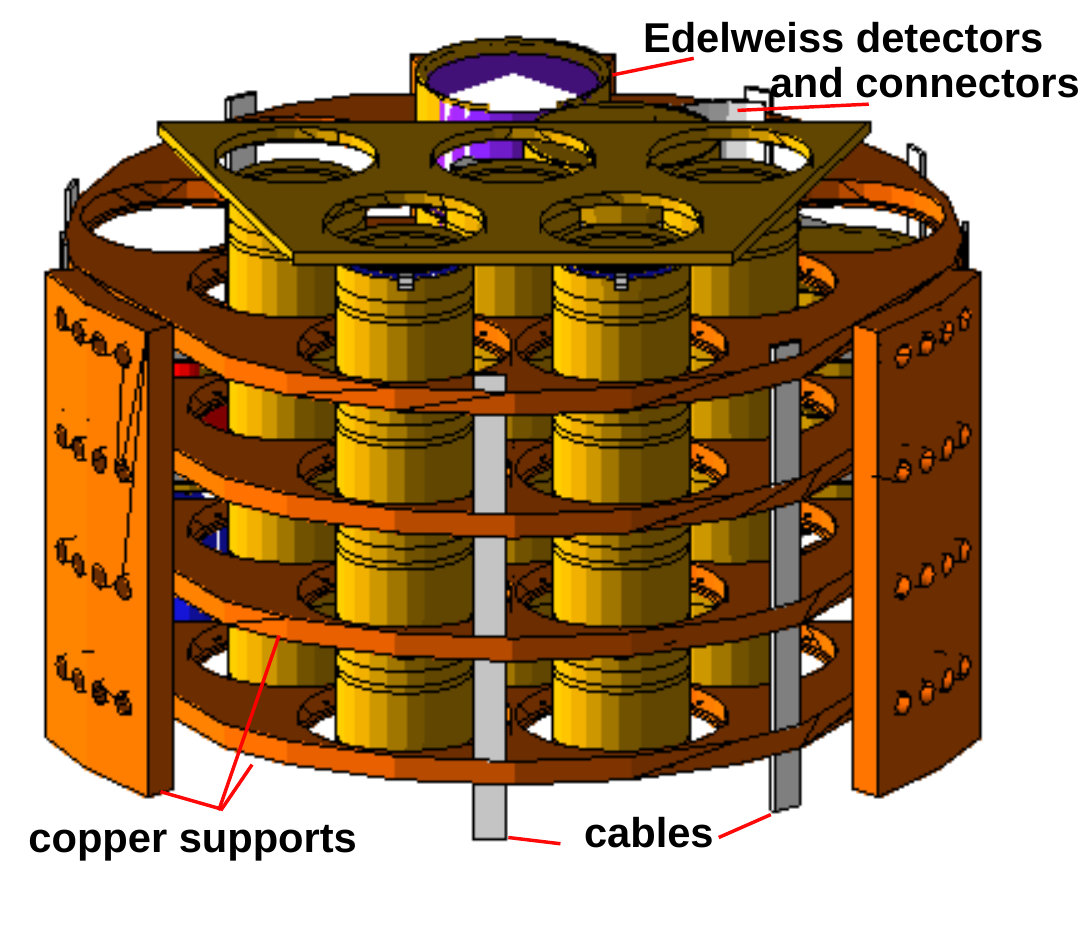}
    \caption{ {\it Left: }The CUPID-Mo experiment in the EDELWEISS cryostat. The five towers on the front contain the CUPID-Mo detectors and the EDELWEISS detectors can be seen behind.  {\it Right: }GEANT4 rendering of the detector chamber in the Monte Carlo simulation geometry. Inside the five towers are placed the LMO crystals, the light detectors, the clamps and the reflective foils (not seen). The readout cables and the structure supporting the towers are indicated.}   
 \label{fig:CUPIDMo-towers}
\end{figure*}
 \\ \indent The array of 20 bolometers is arranged in five towers with four modules each,  as shown in Fig.~\ref{fig:CUPIDMo-towers}. Each tower is suspended by stainless steel springs to mitigate the vibrational noise of the set-up.
The signal from the CUPID-Mo detectors are readout with NOMEX\textregistered~cables, copper and constantan wires on Kapton pads. 
Situated in the same cryostat, the EDELWEISS detectors, visible behind the five CUPID-Mo towers in Fig.~\ref{fig:CUPIDMo-towers}, are equipped with Kapton\textregistered~pads and MillMax\textregistered~ connectors. The detector chamber consists of four copper plates made of NOSV\textregistered~grade copper\footnote{Copper of 99.9975\% purity, produced by Aurubis, Hamburg, Germany.} to support the bolometers, and is able to accommodate 12 detector towers.

\begin{figure}
    \centering
    \includegraphics[width=0.45\textwidth]{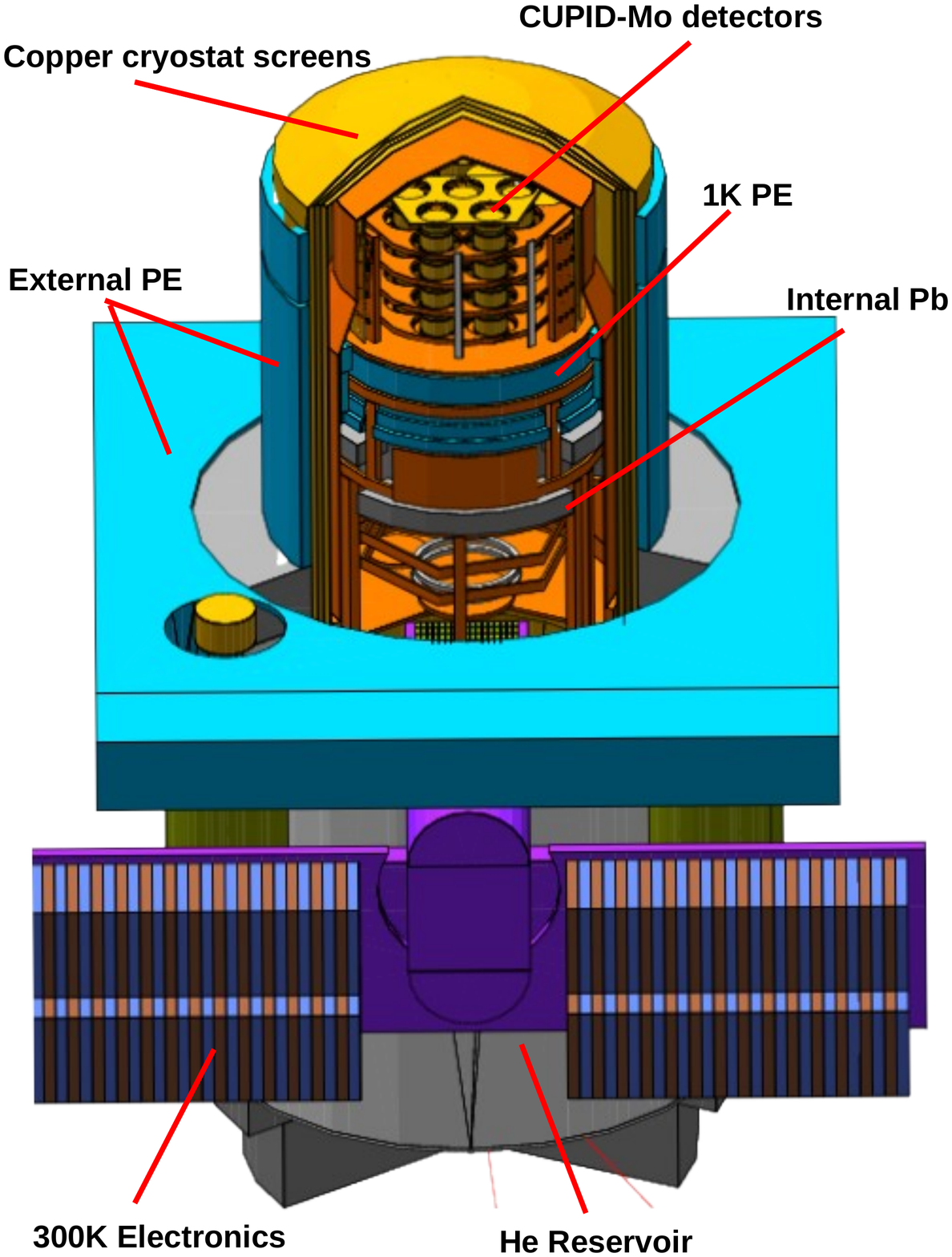}
    \caption{GEANT4 rendering of the CUPID-Mo Monte Carlo simulation geometry, showing the cryogenic set-up.}
    \label{MC_rendering}
\end{figure}

The cryostat involves five thermal copper screens, typically referred to as the 10 mK, 1 K, 50 K, 100 K and 300 K stages respectively. The cryostat screens are made of NOSV and CUC2\footnote{Copper of $>$ 99.990 \% purity} grade copper. An internal polyethylene shield,  to shield against neutrons produced in the set-up components by ($\alpha$,n) reactions or induced by muons \cite{EDWBackgroundpaper}, is mounted between the detectors and the internal lead shield, and
has a temperature of $\sim$1~K.  An internal lead shield of 14~cm Roman lead \cite{Roman_lead} is installed inside the cryostat at 1~K, between the detector chamber and the dilution unit (see Fig. ~\ref{MC_rendering}).  Its main purpose is to shield the detectors from radioactive background of the warm electronics, the cold electronics and the connectors and cables at the 1K stage. 

\begin{figure}
    \centering
     \includegraphics[width=0.5\textwidth]{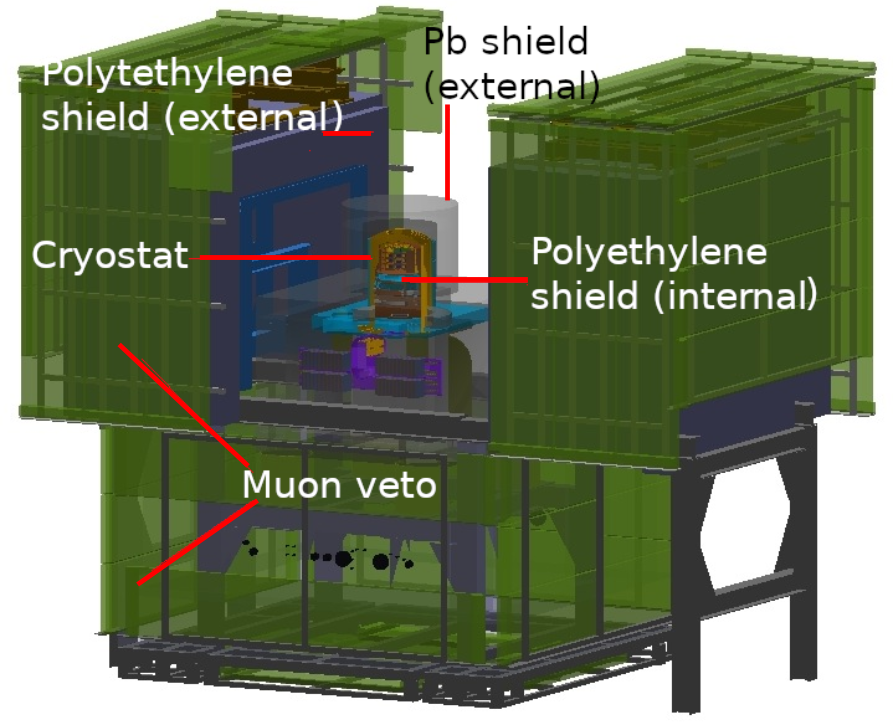}
    \caption{ Visualisation of the EDELWEISS cryostat and shielding as implemented in our MC simulations, we show the cryostat surrounded by the lead shield, the external polyethylene shielding and the muon veto panels. The muon panels are free to move to give a full geometric coverage.}
    \label{fig:shielding}
\end{figure}

The external shielding  closest to the cryostat consists of 20 cm thick lead, with the innermost 2 cm made of Roman lead. The empty space between the lead shield and the outermost thermal screen of the cryostat is flushed with radon depleted air from a radon trapping facility. The average  radon level in the air supplied by the facility   is 20 mBq/m$^3$ \cite{Hodak}. Following the external lead shield, a 50 cm thick polyethylene shield is used to moderate the radiogenic neutron flux.
A plastic scintillator based active muon veto system  surrounds the whole experiment for  muon tagging \cite{EDWMUONVETO} (see Fig. \ref{fig:shielding}).

\subsection{Performances}

CUPID-Mo has shown excellent detector performances in terms of energy resolution ($7.4 \pm 0.4)$ keV FWHM at 3034~keV \cite{Augier:2022} and  $\alpha$ particle rejection $>$ 99.9\% \cite{Armengaud:2019}, demonstrating that the CUPID requirements are within reach. 
Further details on the CUPID-Mo set-up and performances are given in \cite{Armengaud:2019}.

\section{Experimental data}
\label{section:data} The aim of our data processing is to convert the raw data stream into  three calibrated energy spectra: $\beta/\gamma$ like events with energy deposits in a single crystal ($\mathcal{M}_{1,\beta/\gamma}$), events with energy deposits in two crystals ($\mathcal{M}_2$) and of $\alpha$-like events ($\mathcal{M}_{1,\alpha}$). These spectra will then be used in a simultaneous fit to extract radioactive contamination values and describe the observed spectra. The algorithms used for the data processing are described in detail in \cite{Augier:2022} but we will give a summary of the most important steps in the following. We also estimate the detector response parameters (energy resolution, energy bias, efficiencies, light yield) which are needed 
for post-processing the Monte Carlo spectral shapes. 
\subsection{Data taking}
In this paper, we use the same dataset as in  \cite{Augier:2022} with an exposure of $2.71$ kg $\times$ yr of LMO corresponding to $1.47$~kg~$\times$~yr of $^{100}$Mo.
Our data is acquired as a continuous time-stream and digitized at $500$ Hz by the EDELWEISS DAQ \cite{Armengaud:2019} and stored at both CC-IN2P3 (France) and NERSC (USA) for offline analysis.  We acquire runs, periods of around $10-100$ hours of stable data taking, of both physics and calibration data, where a calibration source was placed in the vicinity of the experiment.
We use regular calibrations with a $^{232}$Th/$^{238}$U source to calibrate the  LMO  detectors and a high activity $^{60}$Co source, which generates $^{100}$Mo X-rays in the detectors, to calibrate the LDs. We divide the data into twelve periods of $\sim 1$ month of stable data taking, called {\it datasets}. We discard three short periods of data ($\approx $ 1 week each) due to the inability to accurately calibrate this data. 

\subsection{Data processing}
\label{subsection:data_processing}
\indent We process our data using the C++ softwares {\it Apollo} and {\it Diana} \cite{Alduino:2016, Azzolini2018}, first developed for the CUORE experiment and also used by CUORICINO, CUPID-0 and CUPID   \cite{CUPID:cubic}. 
\\ \indent A complete description of the data processing can be found in \cite{Augier:2022}.
We identify physics events using an optimal trigger, also used for previous CUPID-Mo analysis.
This triggering is used for both the LDs and the LMO bolometers. 
We then store a 3 s waveform for both LDs and the LMO channels for each triggered events.
For each LMO we associate (up to) two light detectors, called  {\it side-channels}, which correspond to the LDs facing this LMO detector. These are numbered S1/S2 where S1 is the LD with the better detector performance (lower noise and higher detector light yield).
\\ \indent We estimate the amplitude of peaks using an optimal filter, which maximises the signal to noise ratio based on inputs of the known signal shape and spectral noise power density. This is done for all LMO events and also the corresponding LD events on the side-channels.
\\ \indent 
Next we correct for thermal gain changes and calibrate our data using dedicated $^{232}$Th/$^{238}$U calibration measurements. 
This calibration is accurate to around $<1$ keV \cite{Augier:2022} which is  sufficient for the binned background model fits.
The LDs are calibrated using the dedicated $^{60}$Co calibrations which produces $\sim 17$ keV Mo X-rays.
\subsection{Multiplicity}
We define coincidences between physics events, where multiple detectors are triggered simultaneously. This provides useful information since events of \onbb decay or \nnbb decay to the ground state are very likely to deposit energy in only one crystal. However, background events in particular from $\gamma$'s are  likely to deposit energy in multiple crystals simultaneously, for example due to Compton scattering in one crystal, or multiple $\gamma$'s from the same decay.
We estimate the multiplicity of an event as the number of pulses in different LMO detectors above our analysis energy threshold (set at 40 keV) within a $\pm$ 10 ms time window. 

\subsection{Data selection}
\label{section:PSD}
Several cuts are used to remove non-physical events (for example noise spikes and cross-talk) or coincidences of two or more pulses generated by events very close in time within the same crystal, called pile-up events. We require that there is only one trigger in the 3~s LMO waveform. We then define a pulse shape discrimination (PSD) cut, described in detail in \cite{Augier:2022}, using a principal component analysis method (PCA). We also define a cut on the pulse rise time and optimal filter based PSD variables\footnote{The optimal filter test values or the $\chi^2$ for rising and falling edges and for the pulse baseline.} which help to cut pile-up like events. Details of the choice of the selection cuts is given in \cite{Augier:2022}.

\subsection{Particle Identification}
\label{section:PID}
Since CUPID-Mo is a dual readout experiment we can discriminate $\alpha$ from $\beta/\gamma$ particles. The use of light detectors also allows us to remove background events in which a particle deposits energy on our LDs. 
We select $\beta/\gamma$ candidate events using the LD signal as following.
We normalise the measured LD signals by defining the variable {\it n}, as the difference between the measured LD energy, $E_{LD}$,  and the mean expected $\beta/\gamma$ LD energy $L$, normalized by the light band width. We compute {\it n}  for $\mathcal{M}_1$ events.
As each LD has different characteristics, the calculation is done for each channel $c$, each dataset $d$ and for both side channels $s$, i.e.:
\begin{equation}
   n_{c,s,d}= \frac{E_{LD}-L_{c,d,s}(E)}{\sigma_{c,d,s}(E)},
\end{equation}
with $E$ the measured LMO energy. The parameter {\it n$_{c,s,d}$} has a distribution expected to be centered at zero for $\beta/\gamma$'s and at a value different from zero for $\alpha$ particles. For details on the determination of the mean expected LD energies and its uncertainty, see \cite{Augier:2022}.

For events with two LDs we expect the 2D distribution of $n_{c,1}$ against $n_{c,2}$ to be a bivariate Gaussian. As we observe no clear correlation we place a radial cut on the variable:
\begin{equation}
    D  = \sqrt{n_{c,1}^2+n_{c,2}^2}.
\end{equation}
If only one LD is available the cut is instead placed just on this {\it n$_{c,s,d}$}. We chose a cut of $D<4$ to select $\beta/\gamma$ events and call this data spectrum $\mathcal{M}_{1,\beta/\gamma}$. We also construct a spectrum of $\mathcal{M}_{1,\alpha}$ events comprised of high energy $\mathcal{M}_1$ events, $E> 3$~MeV, with no light selection applied. 
The same events are obtained with a selection cut  $D>4$, thus, for simplification we have chosen only the energy cut to select $\alpha$ events.
\\ \indent Unlike most other analysis of scintillating bolometers we also develop a light selection cut for \M$_2$ events as described in detail in \cite{ES-paper}. For a \M$_2$ event the scintillation light recorded can be the sum of that from the crystal above and below a given LD. We use the modeling described in \cite{Augier:2022} to compute the expected light detector energy for each physics event accounting for multiple contributions to the light yield.  From this we can define the normalised LD energy for each pulse in a $\mathcal{M}>1$ event. 
 We require that each normalised LD energy (for each channel and side channel) is between $-$10 and 10 $\sigma$, for all but one LD.  In this particular LD we observe an accidental  contamination of $^{60}$Co. Therefore we generally observe $\gamma$ events in the LMO and $\beta$ events in the LD with very large energy compared to scintillation light. For the two LMOs adjacent to this LD we make a cut of $-$10 to 3 $\sigma$, to take into account the energy directly deposited in the light detector. For more details see \cite{ES-paper}. In addition, to further suppress events from this localized $^{60}$Co source we make a {\it global LD anti-coincidence cut} to remove the $\gamma$ background originating from this LD. We remove any events (on non-adjacent LMOs) with a trigger on this LD with energy $>2$ keV within a 5 ms window.

\subsection{Muon veto anti-coincidence}
\label{muveto}
Despite the large rock overburden at LSM, which suppresses most muon events, they still form a possible background source. The EDELWEISS cryostat has a muon-veto system to remove these events, as shown in Fig. \ref{fig:shielding}. We remove events, in each of the \M$_{1,\beta/\gamma}$, \M$_2$ and \M$_{1,\alpha}$ spectra, with a trigger in the veto system within a 5 ms window. With 98\% geometric coverage and the operation voltage adjusted for the aging of the scintillator we expect an $\mathcal{O}$(90\%) tagging efficiency of muons with a minimal impact on the $\beta$/$\gamma$ acceptance \cite{EDWMUONVETO}. 
Since this background was already subdominant and is strongly suppressed by the veto cut we do not include muons in our background model.
\subsection{Delayed coincidences}
\label{section:delayed_coincidences}
Radioactivity from the $^{232}$Th and $^{238}$U decay chains in the LMO crystals could be a significant background in our data. Similar to other analyses of scintillating bolometers \cite{Azzolini:2022, Azzolini:2021}, we can exploit the time correlation of these decay chain events to reduce our experimental backgrounds. In particular, we veto events from the lower part of both chains where there are  backgrounds from $^{214}$Bi ($^{238}$U chain) and $^{208}$Tl ($^{232}$Th chain). For $^{208}$Tl we veto events in $10\times T_{1/2}$ (1830~s) following a suspected $^{212}$Bi $\alpha$-decay. This time window contains $>99.9\%$ of the $^{208}$Tl decays. 
\\ \indent The very low CUPID-Mo radioactivity also enables a novel delayed coincidence cut removing $^{214}$Bi candidate events. The $^{222}$Rn decay chain proceeds as follows:
\begin{align}
        ^{222}\text{Rn} &\xLongrightarrow[\alpha \  5590 \ \ \mathrm{keV}]{3.8\ \mathrm{day}} 
         \mathrm{^{218}\text{Po}} \xLongrightarrow[\alpha \  6115\ \ \mathrm{keV}]{3.1 \ \mathrm{min}} \mathrm{^{214}\text{Pb}} \nonumber \\ &\xLongrightarrow[\beta^{-} \ 1018 \ \mathrm{keV}]{27.1 \ \mathrm{min}}  \mathrm{^{214}\text{Bi}}\xLongrightarrow[\beta^{-} \ 3269 \ \mathrm{keV}]{19.7 \ \mathrm{min}}\mathrm{^{214}\text{Po}}.
\end{align}

We can therefore tag the event based on the $^{222}$Rn or $^{218}$Po $\alpha$ events and a fairly long dead time. We use energy cuts of $5985-6145$ keV for $^{218}$Po and $5460-5620$ keV for $^{222}$Rn to tag $\alpha$ candidates. For either type of $\alpha$ candidate events we then veto events within the same crystal within a time window containing 99\% of events which is evaluated with MC sampling as 13860~s for $^{222}$Rn and $13620$~s for $^{218}$Po. The two possible cuts on $^{222}$Rn or $^{218}$Po improve the rejection power for surface backgrounds.
This cut has a small inefficiency (see section \ref{section:effs}), despite the long veto time.

\subsection{Data spectra}
\label{sec:data_spec}
Based on these cuts we construct the three  data spectra used in our analysis:
\begin{itemize}
\item \M$_{1,\beta/\gamma}$:  Events in one detector identified as $\beta/\gamma$,
\item $\mathcal{M}_2$: Events in coincidence between 2 crystals, the two energies deposited in each crystal are summed,
    \item \M$_{1,\alpha}$: Events in one detector  with alpha energy scale ($>3$~MeV).
\end{itemize}
Because of the relatively fast half-life of $2\nu\beta\beta$ in $^{100}$Mo ($\sim 7 \times 10^{18}$ yr) and extremely low levels of contamination,  relatively few peaks are observed in the $\mathcal{M}_{1,\beta/\gamma}$ spectrum, where the spectrum of $2\nu\beta\beta$ decays of $^{100}$Mo is the dominant feature. The secondary datasets, \M$_2$ and $\mathcal{M}_{1,\alpha}$ however contain a lower fraction of $2\nu\beta\beta$ events and therefore provide useful information to determine the location of radioactive contaminations. 
The experimental spectra after all cuts are shown in Fig. \ref{fig:data}.

\subsection{Data selection efficiencies}
\label{section:effs}
We evaluate the efficiency of our cuts and correct the MC simulations by these values. 
In particular we use events in $\gamma$ peaks from \M$_2$ and \M$_{1,\beta/\gamma}$ spectra to evaluate the efficiency of the PSD (section \ref{section:PSD}), light yield  and rise time cuts (section \ref{section:PID}).
We do not observe that the cuts have any energy dependence in the range of the utilised $\gamma$ peaks (236-2615 keV).
For cuts where the inefficiency can be considered as a dead time, 
the multiplicity, muon veto, delayed coincidence and LD anti-coincidence, we evaluate the efficiency using the $^{210}$Po peak. We evaluate the pile-up efficiency, the probability a pulse will be superimposed with another in a 3~s window, using random noise triggers. 
More details on each of these calculations can be found in \cite{Augier:2022}, and the results are summarised in Table \ref{effs}.
\begin{table}
\caption{Efficiencies for the cuts used on CUPID-Mo data. The PSD and Light Distance cut efficiencies are evaluated using $\gamma$ peaks, and show no energy dependence in the range of the fit.
}
\label{effs}       
\renewcommand{\arraystretch}{1.4}

\begin{tabular}{lll}
\hline\noalign{\smallskip} 
Cut &  Evaluation Method & Efficiency [\%] \\
\noalign{\smallskip}\hline\noalign{\smallskip}
PSD (\M$_{1,\beta/\gamma}$) & \M$_{1,\beta/\gamma}$ $\gamma$-peaks & $95.2\pm 0.5$ \\
PSD (\M$_{2}$) & \M$_{2}$ $\gamma$-peaks & $96.9\pm 0.5$ \\
Light Distance (\M$_{1,\beta/\gamma}$) & \M$_{1,\beta/\gamma}$ $\gamma$-peaks & $99.4\pm 0.4$ \\
Light Distance  (\M$_{2}$) & \M$_{2}$ $\gamma$-peaks & $97.7 \pm 1.8$  \\
Multiplicity & $^{210}$Po  & $99.55\pm 0.07$ \\
Rise time cut  & \M$_{1,\beta/\gamma}$  $\gamma$-peaks & $99.8\pm 0.2$ \\
LD Anti-coincidence & $^{210}$Po  & $99.976 \pm 0.017$ \\
Muon Veto cut & $^{210}$Po  & $99.62\pm 0.07$ \\
Delayed coincidences &  $^{210}$Po       &   $99.16 \pm 0.01$    \\
Pile-up & Noise    &   $95.7 \pm 1.0$ \\ 
\noalign{\smallskip}\hline \hline
Total \M$_{1,\beta/\gamma}$ & & $88.9\pm1.1$ \\
Total $\mathcal{M}_2$ & & $83.3 \pm 2.5$ \\
Total \M$_{1,\alpha}$  & & $94.7 \pm 1.0$ \\ \hline
\end{tabular}
\end{table}
\subsection{Energy scale and resolution}
\label{subsection:energy resolution}
We use the observed $\gamma$ peaks in both background and calibration data to predict the energy linearity and resolution.
Each LMO detector in each dataset has a distinct energy resolution. As in \cite{Armengaud:2021, Augier:2022} we perform a  fit of the 2615 keV peak in calibration data to extract the resolution of each detector-dataset pair. 
We use these resolutions to build a function including a common scale factor $R(E)$ which will be determined for the peaks in background data.
 For our Monte Carlo simulations for each event we sample from a Gaussian with mean $E$ and width $R\times \sigma_{c,d}$, where $\sigma_{c,d}$ is the energy resolution in crystal $c$ and dataset $d$. This energy calibration is discussed in detail in  \cite{Augier:2022}.
\subsection{Features of data spectra}
\label{subsection:peaks}
\begin{figure*}
    \centering
    \includegraphics[width=0.45\textwidth]{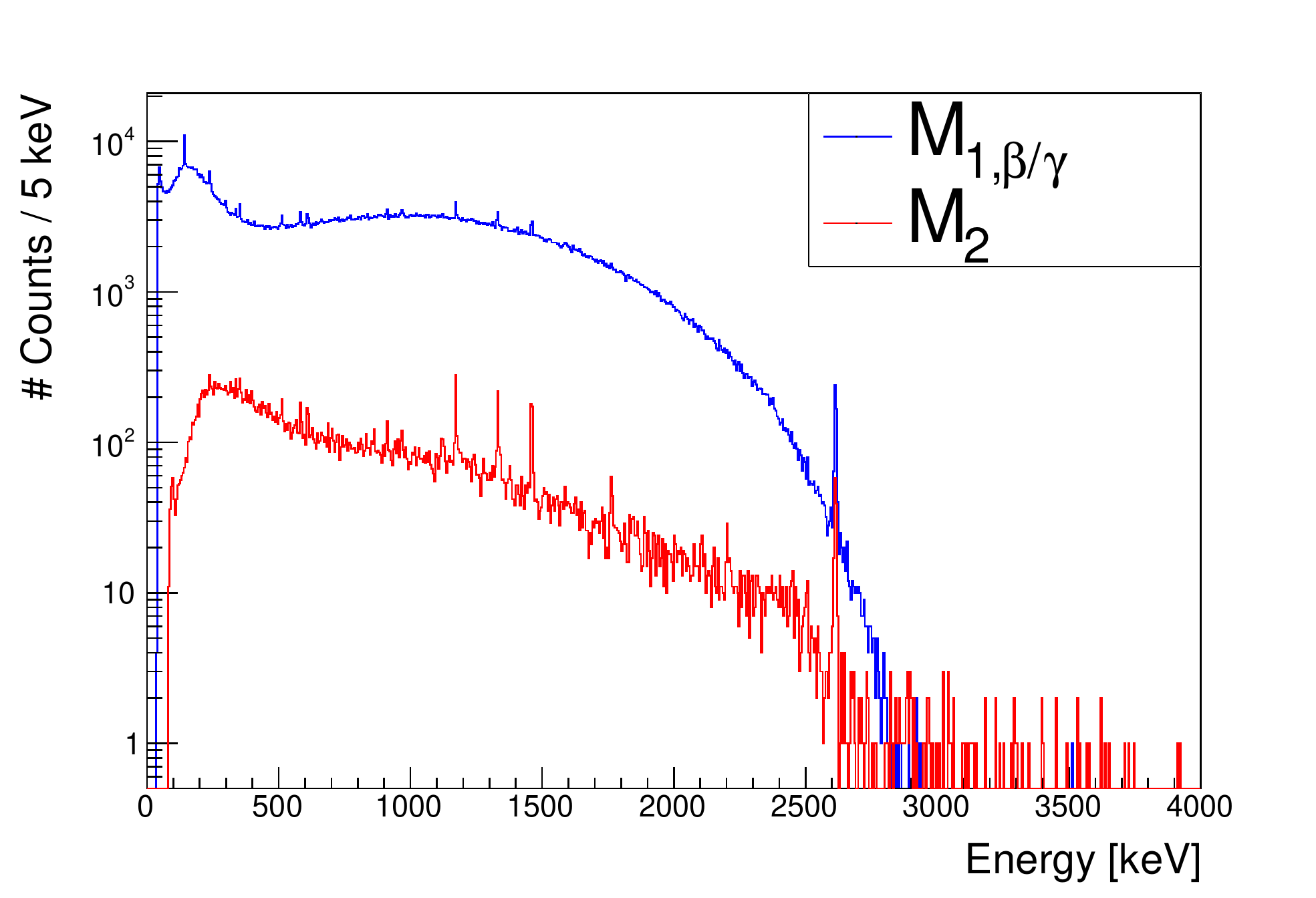} 
    \includegraphics[width=0.45\textwidth]{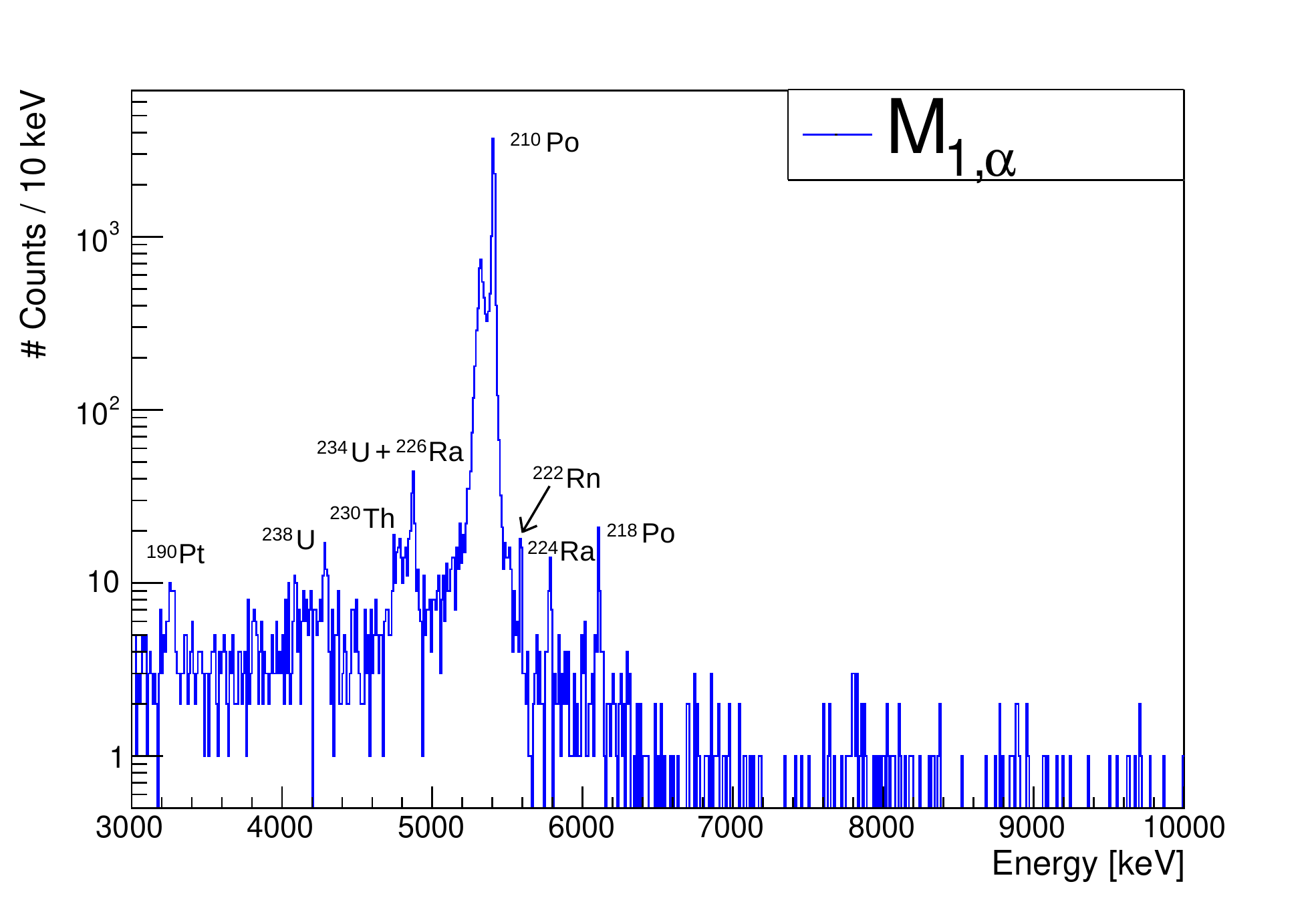}
    \caption{CUPID-Mo experimental data. {\it Left:} \M$_{1,\beta/\gamma}$: events in one detector identified as $\gamma/\beta$. \M$_2$:  events in coincidence between 2 crystals, the two energies deposited in the crystals are summed. {\it Right:} \M$_{1,\alpha}$: events in one detector  with $\alpha$ energy scale.}
 \label{fig:data}
\end{figure*}

 We observe in Fig. \ref{fig:data}  that the spectrum of $2\nu\beta\beta$ decays of $^{100}$Mo dominates the \M$_{1,\beta/\gamma}$ data, whereas the \M$_2$ spectrum has significant contributions from natural radioactivity, shown by prominent $\gamma$ peaks. 
 These consist predominantly of decays from the $^{238}$U and $^{232}$Th decay chains, however we also observe contributions from $^{40}$K and cosmogenic activation products $^{60}$Co and $^{57}$Co. We also observe a short lived peak of $^{99}$Mo, present for $\sim 1 $ dataset, from neutron activation after a calibration with an AmBe neutron source. 
\\ \indent The spectrum \M$_{1,\alpha}$ is dominated by $\alpha$ decays from components very close to the detectors. As shown in Fig.~\ref{fig:data}, in our data we observe a large contribution of $^{210}$Po, $E_{\alpha}=5303$~keV, with both a large $Q$-value and $\alpha$-energy peak and peaks from several other nuclides in the U/Th chains. 
\\ \indent During $\alpha$ decay the energy released is shared between the $\alpha$-particle and recoiling nucleus (NR), with energy $\mathcal{O}$(100~keV). In LMO crystals the range of $\alpha$ particles is about  10~$\mu$m  and a few~nm for nuclear recoils. Therefore we expect to observe a peak at the Q-value of the decay for a LMO bulk event. For surface activity the energy spectrum depends on the implantation depth.  For shallow contribution $\mathcal{O}$(nm) in the crystal the $\alpha$ or recoil could escape, or both could be contained in the crystal. We therefore expect peaks at the NR energy, at the $\alpha$ energy, and at the $Q$-value, with a relatively low flat continuum from partial contained $\alpha$'s or NR. The ratio of these peaks depends on the depth of radioactive contamination.
For a deeper contribution $\mathcal{O}$$(\mu$m) the NR is almost always contained but the $\alpha$'s can still escape after depositing some of its energy, giving rise to a continuum extending from low energies up to the $Q$-value. 
\\ \indent Similarly, for materials facing the crystals we expect a dependence on the implantation depth: at shallow depths the spectrum will be characterised by peaks at the $\alpha$-energy and NR energy, for a deep contribution this will become a flat spectrum from low energy up to the alpha energy.
We note from Fig.~\ref{fig:data} that we generally do not observe clear $\alpha$-energy peaks in our data. However due to the limited statistics the data is still compatible with a full surface contamination. 
The lack of clear $\alpha$-energy peaks creates a challenge for assessing the surface contamination. 

\renewcommand{\arraystretch}{1.4}

\section{Background sources}
\label{section:background_sources}
The background in our experiment is expected mainly from the natural radioactivity in the whole experimental setup, including the detectors. Other contributions from muons, neutrons and environmental gammas are expected to be subdominant, as explained in section \ref{subsection:  others }. To minimize the background, all the materials used to build the experiment have been carefully selected in terms of radiopurity.  To this end,  the daughthers of $^{238}$U and $^{232}$Th decay chains, $^{40}$K, and cosmogenic radionuclides have been measured by High Purity Ge $\gamma$-ray spectroscopy and ICPMS (Inductively Coupled Plasma Mass Spectromery). The CUPID-Mo materials were chosen to minimize the $^{226}$Ra and $^{228}$Th contaminations, as the most critical radioactive backgrounds in the 3 MeV region relevant to \onbb decay searches arise from $^{214}$Bi and $^{208}$Tl decays. Table \ref{cupid-mo-screening} reports the radioactivity in the CUPID-Mo detector components resulting from CUPID-Mo and CUORE measurement campaigns \cite{Alduino:2017}. The materials which are directly facing the crystals (all but the springs from Table~\ref{cupid-mo-screening}) are referred to as {\it close components} in the following. 
The material choice in the \edw cryostat was done to minimize the contaminants at lower energies,  $ \mathcal{O}(100$~keV), which is the region of interest in dark matter searches. Table \ref{edw-screening} shows the radioactivity in the EDELWEISS cryostat materials \footnote{In Table 5, the Kapton connectors, MillMax connectors and Cu Kapton cables belong to the EDELWEISS readout system, while the NOMEX cables are used for the CUPID-Mo readout.}. The NOSV copper is used for the CUPID-Mo detector holders, all the copper parts in the detector chamber and  the cryostat screens (with the exception of the 1~K screen).
\\ \indent We identify the most significant contributions to our  experimental background using the screening measurements and the analysis of experimental data from section \ref{subsection:peaks}. 
\begin{table*}
\caption{Radioactivity values of the components of the CUPID-Mo detectors. All measurements have been made  by ICPMS, with the exception of the Springs, measured by $\gamma$-ray spectroscopy and the surface contamination of the Vikuiti\textsuperscript{\texttrademark} reflective foil, measured with the BiPo-3 detector \cite{BiPo}. }

\label{cupid-mo-screening} 
\begin{tabular}{llllll}
\hline\noalign{\smallskip} 
 Element     &    Total mass [g]          &  \multicolumn{3}{l}{  Activity [mBq/kg]} &  \\
                                      &                         &    { $^{238}$U}   & { $^{232}$Th}    &  Others       \\      \noalign{\smallskip}\hline           
  Ge-LD \cite{Azzolini:2019}   &   $27.4$                    &  $<1.9\times 10^{-2}$ &  $<6 \times 10^{-3}$   &       \\
 NTD \cite{Alduino:2017}     &   $2$                       &  $< 12$    &   $<4.1$     &     \\ 
PTFE clamps \cite{Alduino:2017} &  $216$                  &  $< 2.2\times 10^{-2}$           &   $<6.1\times10^{-3}$    &       \\
Reflectors (Vikuiti\textsuperscript{\texttrademark})&  $10.08$   & $(1.7 \pm 0.5)\times 10^{-1}$      & $(4.9 \pm 1.2)\times 10^{-2}$  &   $^{214}$Bi :  $(1.0  \pm 0.4)$ nBq/cm$^{2}$ \\
    {Springs}           & 8.1          &  $^{226}$Ra: $11  \pm 3$  &  $^{228}$Th: $21  \pm 5$      &     $^{228}$Ra: $26 \pm 9$;  $^{40}$K: $(3.6 \pm 0.4)\times 10^3$        \\ 
\end{tabular}

\end{table*}

\begin{table*}
\caption{Radioactivity of the components in the EDELWEISS setup. All measurements have been made  by HPGe $\gamma$-spectroscopy. The MillMax connectors have also been measured by ICPMS.}
\label{edw-screening} 
\begin{tabular}{lllll}
\hline\noalign{\smallskip} 
Element     &    Mass [g]       &  \multicolumn{3}{l}{  Activity [mBq/kg]}   \\
            &                     &    { $^{226}$Ra}   & { $^{228}$Th}    &  Others       \\      \noalign{\smallskip}\hline           
     {Kapton connectors} &  $33.12$            & $14 \pm 7$       & $67 \pm 31$    &     \\
   {Cu Kapton cables }& $510$         & $8 \pm 6$     & $15 \pm 10$   &       \\
  NOMEX cables & $4$         &    $21$     & $19$      &       \\
        MillMax connectors   & $0.5$          &  $(1.0 \pm 0.6)\times 10^2$             & $(1.0 \pm 0.2)\times 10^3$      &   $^{238}$U: $(1.2 \pm 0.2)\times 10^4$         \\
  {Brass screws} & $2000$         &   -      & $3.5\pm 0.9$   &   $^{210}$Pb: $(6 \pm 3)\times10^2$; $^{137}$Cs: $2.6 \pm 1.5$   \\
  Cu NOSV$^a$    & $2.89\times 10^5$      &   $<0.040$                    & $(2.4 \pm1.2)\times 10^{-2}$     &         \\
 Cu CUC2$^a$    & $6.5\times 10^5$       &          $(2.5  \pm 1.5)\times 10^{-2}$          &   $(3.3\pm 1.6)\times 10^{-2}$     &       \\
  {PE internal }  & $1.51\times 10^5$      &          $0.65  \pm 0.08$          &   $0.30  \pm 0.07$  &     \\
  {Conn. 1K to 100K }  & $430$       &          $(2.6  \pm 0.4)\times 10^3$          &   $450  \pm 44$    &       \\\hline
\end{tabular}

$^a$ M. Laubenstein, private communication
\end{table*}
We can broadly categorise our background sources into four groups:
\begin{itemize}
    \item \emph{Close source}: Radioactivity in the LMO crystal, reflective foils, LDs, PTFE clamps and NTDs, directly facing the crystals;
    \item \emph{10 mK source}: Sources of activity in the 10 mK stage of the cryostat but not directly facing the LMO crystals (springs, cables, connectors, copper plates for bolometer support), as shown in  Fig.~{\ref{fig:CUPIDMo-towers}};
   \item \emph{Infrastructure source}: The copper cryostat screens and the internal shieldings, see Fig.~{\ref{MC_rendering}};
    \item \emph{External}: Activity originating from outside the 300 K Cu shield. 
\end{itemize}

\subsection{Other contributions - muons, neutrons and environmental gammas}
\label{subsection:  others } 
The muon flux at the LSM is 5 muons/m$^2$/day \cite{EDWMUONVETO}. Muons would generally deposit energy in multiple detectors and be strongly suppressed by anti-coincidence with the muon veto detector (see section \ref{muveto}), therefore we do not include them in the background model. \\ \indent
Neutrons may induce background in the  $0\nu\beta\beta$ region of interest (ROI) if they are captured in the materials of the setup, producing high energy gammas.  
The thermal neutron flux in LSM has been measured as  $(3.6\pm0.05~\text{(stat.)}\pm0.27~\text{(syst.)})\times 10^{-6}$~neutrons/s/cm$^2$ \cite{Evgeny_neutrons_lsm} and the ambient neutron flux (fast plus thermal) has been estimated $\sim$10$^{-5}$ neutrons/s/cm$^2$ in \cite{Evgeny_neutrons_lsm,Fiorucci_thesis}.
Previous work \cite{Lemrani} showed that 48 cm of polyethylene reduces the neutron flux by a factor  2 $\times$ 10$^6$. Taking into account the surface of the CUPID-Mo detectors, we get that the neutron flux expected is less than 1 neutron/year. Thus, ambient neutrons are not taken into account among our background sources. 
\\ \indent
The gamma flux at LSM has been measured with a portable Ge detector at several locations in the laboratory. At the place where the EDELWEISS set-up is installed, the flux of 2.6 MeV photons was measured as $5.1 \pm 0.2 \ \text{(stat.)}\times 10^{-2}$ $\gamma$/s/cm$^2$ \cite{Gurriaran}. Considering that 20 cm of lead reduce the flux by about a factor 10$^4$, then the contribution of environmental gammas may not be negligible. We expect about 6 photons of 2.6 MeV on the detectors surface during the course of all data taking. We take them into account by generating decays at the level of the outermost cryogenic thermal shield, as the spectral shapes measured in the detector from a source generated outside the external lead and outside the outermost cryogenic thermal differ slightly only below 500~keV.  

\section{Monte Carlo simulations}
\label{section:MC}
The Monte Carlo simulation is developed in GEANT4 and implemented with version 10.04 \cite{Geant4}. The MC simulation program developed by the EDELWEISS collaboration \cite{EDWBackgroundpaper}, has been adapted to include the CUPID-Mo detectors and to include the features described below.  \\ \indent
We sample the \nnbb decay spectra of $^{100}$Mo and we consider both the HSD and SSD mechanisms with their parametrization from \cite{Kotila1, Kotila2}. 
\\ \indent The radioactive decays in the  components of the experiment are generated using both Decay0 \cite{decay0} and GEANT4. For decay chains in close sources we use the GEANT4 class G4RadioactiveDecay. This allows to generate sub-chains, for example $^{226}$Ra to $^{210}$Pb.  We store the final position of the nuclear recoil, and use it as the initial condition for the next decay, along with the time difference.  This allows for the accurate consideration of pile-up from events out of the same decay chain and of the delayed coincidence cuts (see section \ref{section:delayed_coincidences}).
From the simulations we then store the energy deposited in both the LMO and LDs. For the $^{232}$Th decay chain, we generate separately $^{232}$Th, $^{228}$Ra to $^{228}$Th and $^{228}$Th to $^{208}$Pb, since $^{228}$Ra and $^{228}$Th have long half-life and so secular equilibrium cannot be assumed. Similarly for the $^{238}$U chain we generate separately $^{238}$U to $^{234}$U, $^{234}$U, $^{230}$Th, $^{226}$Ra to $^{210}$Pb and $^{210}$Pb to $^{206}$Pb.
\\ \indent We use Decay0 for most external sources, not directly facing the crystals, where pile-up events in the same crystal from subsequent decays in a chain are unlikely, and delayed coincidence cuts through the tagging of $\alpha$ events is impossible.
For the $^{238}$U decay chain we generate the $\beta$ emitters $^{214}$Pb and $^{214}$Bi. Since they are in secular equilibrium, we combine their spectra to reduce the  number of components in the background model fit. We also generate in some components $^{210}$Bi which is not assumed to be in equilibrium. For the $^{232}$Th decay chain we generate  $^{212}$Pb, $^{212}$Bi and $^{208}$Tl out of the $^{228}$Th sub-chain and combine them into one spectrum. We also generate $^{228}$Ac which is not assumed to be in equilibrium.
 \\ \indent In addition to the $^{238}$U and $^{232}$Th chains, we simulate $^{40}$K contamination in the springs and the outermost cryogenic thermal shield. We have also considered  $^{60}$Co from cosmogenic activation in copper as well as $^{87}$Rb and $^{90}$Sr+$^{90}$Y in the crystals. A $^{99}$Mo contribution due to neutron activation in the first days of data taking is also simulated in the crystals.
 \\ \indent The decays are generated in the bulk of the components and also the surface for close sources, 
 where surface contaminants can produce a distinct energy spectrum compared to bulk contamination. Surface contaminations are modelled with an exponential density profile e$^{-x/\lambda}$, where $\lambda$ is a variable depth parameter. 
 \\ \indent
 The particles are propagated through the experimental geometry using the Livermore low energy physics models \cite{Livermore_Phys_List}.  We use production cut lengths\footnote{Production cuts apply to the production of secondaries. Below the cut, the primary is tracked down to zero energy using a continuous energy loss.} of 1~$\mu$m for $e^{-}/e^{+}$ and 10 $\mu$m  for $\gamma$'s. 
 For LMO these correspond to  1 keV energy thresholds for both $e^{-}/e^{+}$ and $\gamma$'s. This choice is based on a study of the impact of the production thresholds on the detected spectra. Thresholds  of 1 keV and 250 eV for LMO give comparable spectra, while  the computing time is significantly reduced for a 1 keV cut length. 
 \subsection{Geometry}
\label{subsection:MC_geometry}
We implement a detailed geometry of the CUPID-Mo towers in the MC simulations. In particular, we reproduce the size of each LMO crystal \cite{Armengaud:2019} on an individual basis to take into account variations between the crystals. We also include the Ge LDs, the PTFE clamps, the reflective foils surrounding the crystals and the copper holders which are implemented as accurately as possible.
 Fig. \ref{fig:CUPIDMo-towers} shows the GEANT4 rendering of the simulation geometry of the 10 mK chamber, with the five towers of CUPID-Mo in the front.  We  included the readout cables, the  springs, the EDELWEISS Ge detectors and their connectors.  The copper structure supporting the crystals, composed of four copper plates and three copper columns made of NOSV copper, is also included in the simulated geometry. The four copper plates are held by brass screws with a relatively high mass  (see Table~\ref{edw-screening}) which have  been modeled as well. 
 \\ \indent Fig. \ref{MC_rendering} shows the simulated geometry of cryostat and electronics. The 10 mK, 1 K, 50 K, 100 K and 300~K thermal screens are included individually. The  internal polytehylene shielding and lead shielding are also implemented in the geometry. We also note that the geometry extends and includes far components below the 1~K lead shield that are less radiopure, like the dilution unit, the 300~K electronics, the pumps and the He reservoir that is expected to be important for neutron simulations. 
 
 \subsection{Detector response model}
To compare the simulated spectra to the measured data we need to account for the finite energy resolution and response of the detectors. The following features are accounted for through a post processing of the MC simulation spectra:
\begin{itemize}
    \item Energy resolution;
    \item Energy threshold of 40 keV;
    \item Event multiplicity;
    \item Scintillation light and LD resolution;
    \item Cut efficiencies;
    \item Inactive periods of detectors;
    \item Pile-up and delayed coincidences in decay chains.
\end{itemize}

We compute the energy resolution per detector-dataset pair as explained in section \ref{subsection:energy resolution}.
In particular, for a pulse with energy $E_{MC}$ in channel {\it c} and dataset {\it d} we sample from a Gaussian with mean $E_{MC}$ and standard deviation $R\times \sigma_{c,d}$.
As is done in experimental data we discard pulses below the energy threshold, $<40$ keV, and compute the multiplicity as the number of detectors with $E>40 $ keV for each simulated event. 
 \\ \indent We also reproduce the signals measured in the light detectors. We have parameterised the scintillation light energy measured by the LD in data as a function of LMO energy as a second order polynomial, for each LMO and side LD channel. We also parameterise the LD energy resolution as a Gaussian with standard deviation $\sqrt{p_0^2+p_1E}$. We use this parameterisation to generate a random scintillation light yield for each event which is summed with the energy deposited in MC from direct particle interactions.
We use these light detector energies to reproduce the light yield cuts in the same way as in experimental data in section \ref{section:data}.
\\ \indent
To account for inactive detectors we assign a random timestamp from the data taking period to each simulated MC event. This allows us to apply the same cuts to the simulated data and remove events from detectors considered inactive and to account for the reduction of event multiplicity in these periods.

 \subsection{Simulated background sources}
\label{subsection:MC sources}
Some components produce indistinguishable spectra of energy deposits in the crystal, or in other words, they exhibit degenerate spectral shapes. In this case, we either group them, or generate the radioactive decays in only one, which accounts for all elements with degenerate spectra.  This simplification reduces the number of free parameters in the fit of the simulations to the data, however, we need to keep in mind that the posterior distributions account for the sum of the  grouped elements. 
\\ \indent The reflective foils, the PTFE clamps, 
and all other passive elements directly facing the crystals produce degenerate spectra. For this reason we have generated radioactive decays only in the {\it Reflectors}. 
We  simulated radioactive decays separately in the connectors, the cables and the springs in the detector chamber at 10 mK shown in Fig. \ref{fig:CUPIDMo-towers}, right. 
All the copper elements made of NOSV at the 10 mK stage (holders, four plates, three columns and 10 mK cryostat screen) have been grouped in one background source, which we refer to as {\it Copper supports}. \\
 \indent We have found that all thermal screens exhibit degenerate spectra, thus we group the screens made of NOSV copper and refer to as {\it Cryostat screens}. 
 We have also found that the  {\it internal polyethylene shielding} spectrum is degenerate with the one from internal lead shielding. Thus, we have chosen to include only the internal polyethylene shielding contribution in the fit. This element takes into account all contributions from background sources below the internal lead shielding, as the elements of the dilution unit or the  300~K electronics also shown in Fig. \ref{MC_rendering}.



In addition we have considered as a source the outer cryostat screen, called {\it screen 300 K}. This volume also includes the contribution  from  radon present in the air between the 300 K screen and the external lead shielding. 
\section{Background model}
The goal of the background model is to describe the data (section \ref{section:data}) with the MC simulations (section \ref{section:MC}). The parameters of the model then tell us the radioactive contamination of various components of the experiment. We use a binned simultaneous maximum likelihood fit, performed in a Bayesian framework with a Markov Chain Monte Carlo (MCMC) approach \cite{MCMC}, developed by  CUORE and further optimized by CUPID-0 collaborations \cite{Alduino:2017, Azzolini:2019} using the JAGS software \cite{JAGS_1, JAGS_2}.  We model the data in spectra $i$ (\M$_{1,\beta/\gamma}$,  \M$_2$ 
and \M$_{1,\alpha}$) and energy bin $b$ as:
\begin{equation}
    f_i(E_b;\vec{N})=\sum_{j=1}^{N_s} N_{j}\times f_{j,i}(E_b).
\end{equation}
 The sum $j$ runs over the simulated MC sources, $N_{j}$ is a scaling factor for each source (shared by all three spectra) and $f_{j,i}(E_b)$ are the simulated MC spectral shapes, with $E_b$  the energy of bin {\it b}. The likelihood function for data  $\mathcal{D}$, including the 3 spectra \M$_{1,\beta/\gamma}$,  \M$_2$ and \M$_{1,\alpha}$  is then given by the product of Poisson distributions, $\text{Poiss}(n_{i,b};f(E_i(E_b;\vec{N})))$, for $n_{i,b}$ observed counts in bin $b$ of spectrum $i$, and prediction $f_i(E_b;\vec{N})$ for the set of parameters $\vec{N}$:

\begin{flalign}
\nonumber
    &\text{ln}\mathcal{L}\big(\mathcal D|f(E_b;\vec{N})\big)=
     \sum_{i=1}^{3} \sum_{b=1}^{N_b(i)} \ln{(\text{Poiss}(n_{i,b};f_i(E_b;\vec{N})))} \\ &=  
     \sum_{i=1}^{3} \sum_{b}^{N_b(i)} n_{i,b} \times  \ln{(f_i(E_b;\vec{N}))} 
   - f_i(E_b;\vec{N}) - \ln(n_{i,b}!).
\label{likelihood}
\end{flalign}
Here the sum $i$ is over the three data spectra and $b$ goes over the bins in each spectrum.

JAGS samples the full posterior probability distribution $p(\vec{N}|\mathcal D)$ given by Bayes theorem:
\begin{equation}
    p(\vec{N}|\mathcal D) = \frac{\mathcal{L}(\mathcal D|\vec{N})\times \pi(\vec{N})}{p(\mathcal D)},
\end{equation}
using MCMC.
The prior probabilities, $\pi(\vec{N})$ are discussed in section \ref{section:priors}.  
For each parameter we also extract the marginalised posterior distribution by integrating over the parameter space $\Omega$ (excluding the parameter of interest):
\begin{equation}
    p(N_j|\mathcal D)=\int_{\Omega} p(\vec{N}|\mathcal D)d\vec{\Omega}.
\end{equation}
We choose the mode of the marginalised distribution as our point estimate of the parameter and we compute, by integrating, the smallest 68\% Bayesian credible intervals, c.i., around the mode. If the lowest 68\%  includes the value zero, we give an upper limit at 90\% c.i. 




\subsection{MC simulation of $^{56}$Co calibration source}
\label{subsection:  56Co } 
We have performed a calibration with a $^{56}$Co source to validate the energy calibration and resolution of the CUPID-Mo detectors
in the \onbb ROI, at $\sim$3 MeV. The measurement is also useful to test and validate the implementation of the Monte Carlo simulations.
Two $^{56}$Co sources with an activity of $41 \pm 8$ Bq, measured with HPGe $\gamma$ spectroscopy inmediately after the calibration, were placed on the outer cryostat screen, inside the external shielding. The configuration was chosen to achieve the highest uniform counting rate in the ROI for all detectors with a total rate below 0.125 Hz as an upper limit on the tolerable pile-up.
\\ \indent We have performed a fit of the calibration data to a MC simulation of the $^{56}$Co sources summed with a background component (detailed later in this section) and pile-up, with only uniform priors. We describe in section \ref{subsection:random} how the spectral shape of the pile-up events is obtained. 
The fit has thus three parameters: the normalization factor of the background, the one of the $^{56}$Co sources, and the one of the pile-up events. 
We know the normalization of the background from the background model fit. Comparing it to the normalization factor of the background in the calibration data, we obtain the efficiency of the cuts in the calibration data (68.7$\pm$1.4)\%.   
 From the normalization factor of the $^{56}$Co sources we obtain the activity of the sources without the efficiency correction. Using the estimated efficiency, we derive the final activity of the $^{56}$Co source of ($50 \pm 1$)~Bq, which is in good agreement with the measured activities.
 
\begin{figure*}
\centering
\includegraphics[width=1\textwidth]{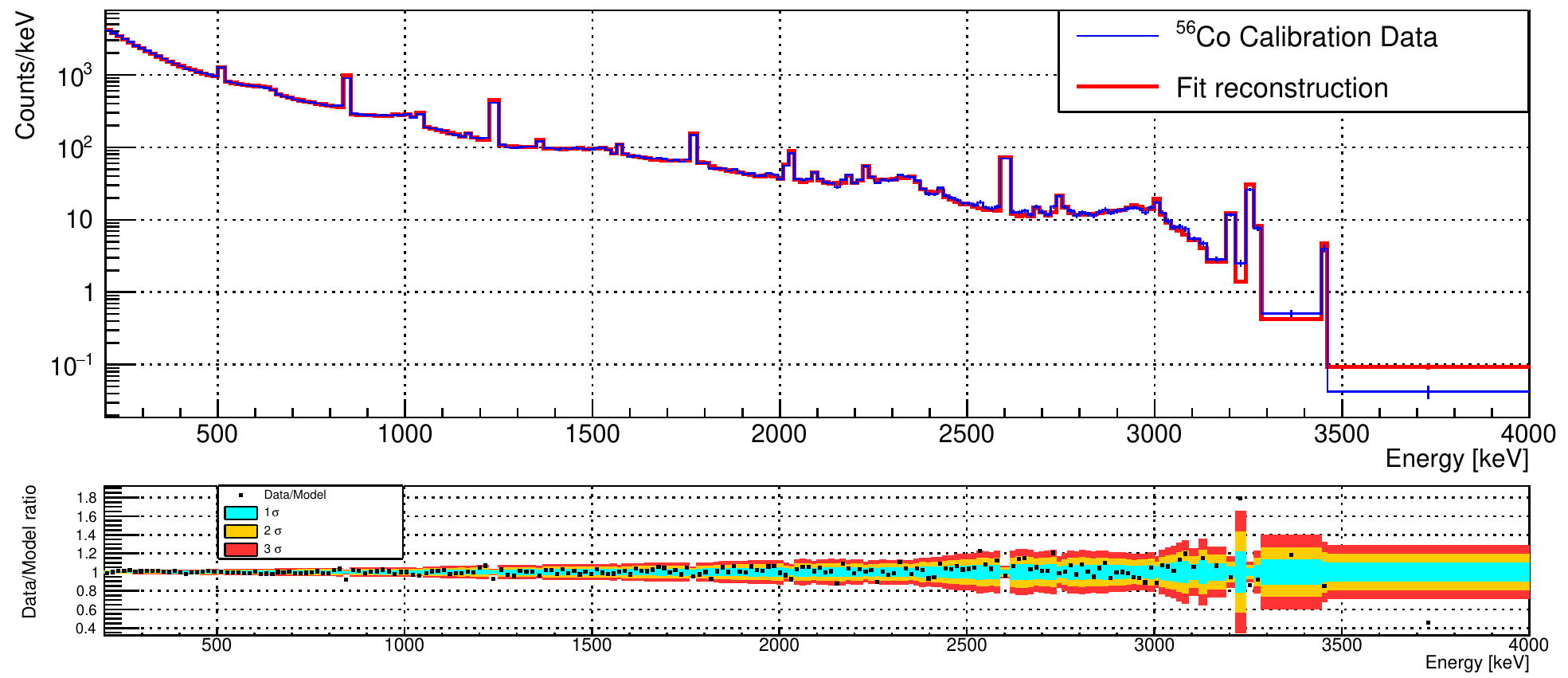}
\caption{Top: Comparison between the $^{56}$Co  calibration data and MC simulations, \M$_1$ data, with a variable binning in the region between 200~--~4000~keV.  Bottom: Bin by bin ratio of data and MC. Most of the values are within 3 $\sigma$, with discrepancies  below 20\%. }
\label{fig:Co56_5keV}
\end{figure*}

\begin{figure}
    \centering
    \includegraphics[width=0.45\textwidth]{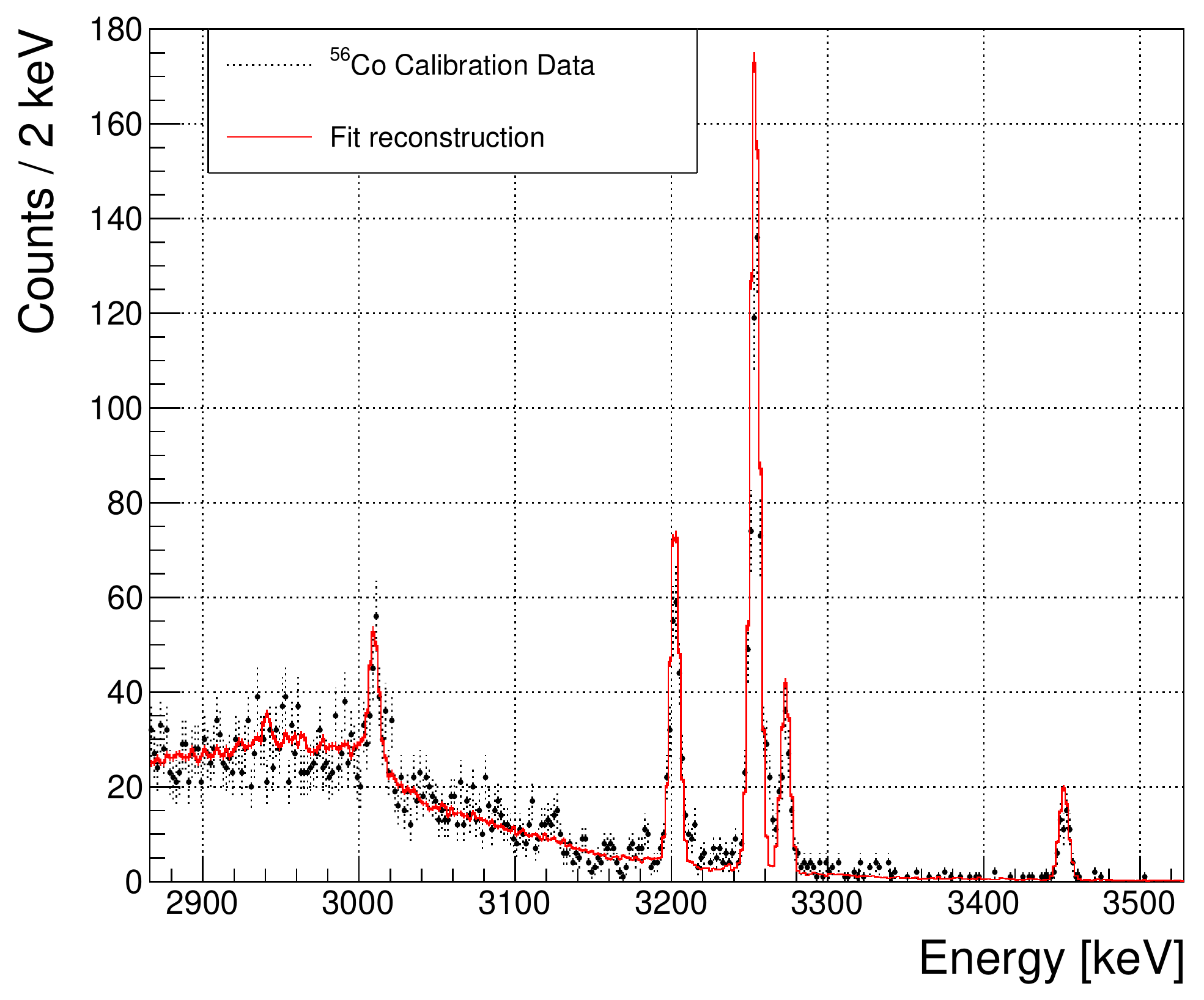}
    \caption{Comparison between the $^{56}$Co  calibration data and MC simulations, \M$_1$ data in the region of interest.}
    \label{calibration_zoom_comparison}
\end{figure}

The model shows good agreement with the data in the whole energy range of the fit 200~--~4000~keV, as shown in Fig. \ref{fig:Co56_5keV}. 
In Fig. {\ref{calibration_zoom_comparison}} we present the region above 2800 keV with 2 keV binning, where the comparison in this region can be better appreciated. This fit  shows that the MC implementation of the set-up is accurate and that the MC is able to describe well  the  data. 

\subsection{Background sources list}
The background model fit includes 41 background sources associated to the bulk volume of the components identified in section \ref{subsection:MC sources}. We included 2 additional sources of surface contamination: the LMO crystal and the Reflective foil (representing all sources facing the crystals).
For a detailed list of the sources associated to radioactive contaminants we refer to Tables~\ref{crystal}  and \ref{tab:Fit_result}.
The complete list of the background sources in our fit is:
\begin{itemize}
\item Crystal:
\begin{itemize}
    \item \nnbb decay of $^{100}$Mo to $^{100}$Ru ground state,
    \item $2\nu \beta \beta$ decay of $^{100}$Mo to $^{100}$Ru 0$_1^+$   excited state,
    \item pile-up (random coincidence of 2 events in the same crystal happening so close in time that the signal is equivalent to that of the sum of the two events),
    \item $^{99}$Mo,
    \item 12 bulk sources of natural radioactivity detailed in Table~\ref{crystal},
    \item 8 sources associated to surface contamination, listed in Table~\ref{crystal},
       \item $^{210}$Pb surface contamination with 1 nm and 1 $\mu$m implantation depth.
\end{itemize}
\item Reflectors:
\begin{itemize}
 \item 3 sources associated to bulk contaminations, Table \ref{tab:Fit_result},
 \item 8 sources associated to surface contaminations, Table \ref{tab:Fit_result},
 \item $^{210}$Pb surface contamination with 100 nm and 1 $\mu$m implantation depth.
 \end{itemize}
 \item Close sources, 10 mK and infrastructure: 27 sources associated to the bulk volume of these components, listed in Table~\ref{tab:Fit_result},
 \item Random coincidence of 2 events in two different crystals called accidentals.
\end{itemize}

A total of 67 sources are included in the fit. 
As mentioned in section \ref{section:MC} we have modelled surface contaminations with an exponential density profile e$^{-x/\lambda}$. We have simulated surface contaminations with $\lambda$ = 10 nm and 10 $\mu$m for all radionuclides in the U and Th chains. The choice 10~nm is driven by the fact the typical range of recoiling nuclei is of the order of some nm for $\alpha$ decays in the U and Th chains. The choice of  10~$\mu$m considers that mechanical crystal preparation including cutting, cleaning and polishing can lead to deep surface damage and implantation depths of $\sim \mu$m, however depths $>$ 10 $\mu$m  would be effectively equivalent to bulk contaminations. We observed that the component corresponding to the shallow surface contamination of the crystal, with $\lambda = 10$~nm, is needed to properly  fit our data. Surface contaminations with $\lambda = 10 \ \mu$m give activities which are compatible with zero. Thus, for simplification to minimize the number of degrees of freedom, we choose to include in the fit only the crystal surface contaminations with  $\lambda$ = 10 nm implantation depth. Due to the small thickness (70 $\mu$m) and low density of the Reflectors, 
surface contaminations with $\lambda$ = 10 $\mu$m are degenerate with bulk contaminations. Both produce continuous spectra due to the partial energy loss of $\alpha$ particles in the Reflector and the detection of the remaining kinetic energy in the crystal. We have therefore 
chosen to include only the surface contaminations with $\lambda$~=~10~nm for the Reflectors in the background model fit.
\\ \indent To simulate surface and bulk contaminations in the crystal and the Reflectors, we have generated the decay chains to take into account time correlations and exploit the delay coincidences (see section \ref{section:MC}), as done for the data.
We observed that the bulk contamination in the Reflectors produce a flat spectra independent of the specific part of the radioactive decay chain and our fits showed the activities of the various subchains (excluding $^{210}$Pb to $^{206}$Pb) were compatible. We hence simplify the fit model by assuming that the entirety of the U/Th (excluding $^{210}$Pb to $^{206}$Pb) chain is in secular equilibrium for the Reflectors. 
\\ \indent In addition to the U/Th chains, other contaminations have been included in the crystals. In particular, $^{40}$K  can be found as a result of an initial contamination of the powder used to grow the crystals \cite{Armengaud:2017}. Some anthropogenic radionuclides due to fall out can also be found in enriched crystals \cite{Oksana_Polishchuk}, thus, we have included the bulk contaminations  $^{87}$Rb and  $^{90}$Sr+$^{90}$Y, which are pure $\beta$-emitters. $^{99}$Mo was produced by activation during a neutron calibration with an AmBe source.  
 \\ \indent For all sources that aren't facing the crystals we have simulated decays of the daughters of $^{226}$Ra and $^{228}$Th. 
We identify in Table \ref{edw-screening} a large content of $^{210}$Pb in the brass screws holding the detector plates. We have simulated $^{210}$Bi in this component and use it to account for this contamination in all 10 mK and infrastructure sources. 
Cobalt isotopes are expected to be primarily the result of cosmogenic activation in copper. 
We have therefore chosen to locate $^{60}$Co and $^{57}$Co in the {\it Copper supports} and use it to account for this contamination in all 10~mK and infrastructure sources. 
In all the components we use $^{228}$Ac $\gamma$ emitter with three main $\gamma$ peaks clearly visible in the data, not in equilibrium with $^{228}$Th. In doing so, we have observed that for the 300~K screen the two values from the fit are compatible with equilibrium thus, to reduce the number of components in the fit we have combined $^{228}$Ac and $^{228}$Th.

\subsection{Choice of priors}
\label{section:priors}
We consider informative priors both from screening measurements (section \ref{section:background_sources}) and from other independent measurements. We have informative priors on the contribution of the:
\begin{itemize}
    \item $2\nu \beta \beta$ $^{100}$Mo to $^{100}$Ru 0$_1^+$   excited state, which has been taken as $T_{1/2}=(6.7 \pm 0.5) \times 10^{20}$ yr  (average value from \cite{Barabash:2b2n}),
    \item Stainless steel springs included in the set-up to mitigate vibrational noise. These are modelled with high accuracy in the MC, and the values measured by HPGe and used as priors are given in Table \ref{cupid-mo-screening},
    \item Random coincidence  (pile-up and accidentals) events, determined from the rate of single events and  from a measurement with a calibration source (see below).
\end{itemize}
\subsubsection{Random coincidence events}
\label{subsection:random}
Energy deposition in either one or two crystals randomly coinciding in time can cause non-negligible contributions to the high energy region both in \M$_{1,\beta/\gamma}$ and even more so in $\mathcal{M}_2$. This is a particular concern for $^{100}$Mo, due to the fast rate of $2\nu\beta\beta$ decay of $T_{1/2} \sim 7 \times 10^{18}$ yr \cite{Barabash:2020}, or around $2$~mHz per detector. The events in two different detectors are referred to as accidentals and contribute to $\mathcal{M}_2$ spectrum. The random coincidences in the same detector are referred to as pile-up and contribute to the  $\mathcal{M}_1$ spectrum.

\begin{figure}
    \centering
    \includegraphics[width=0.45\textwidth]{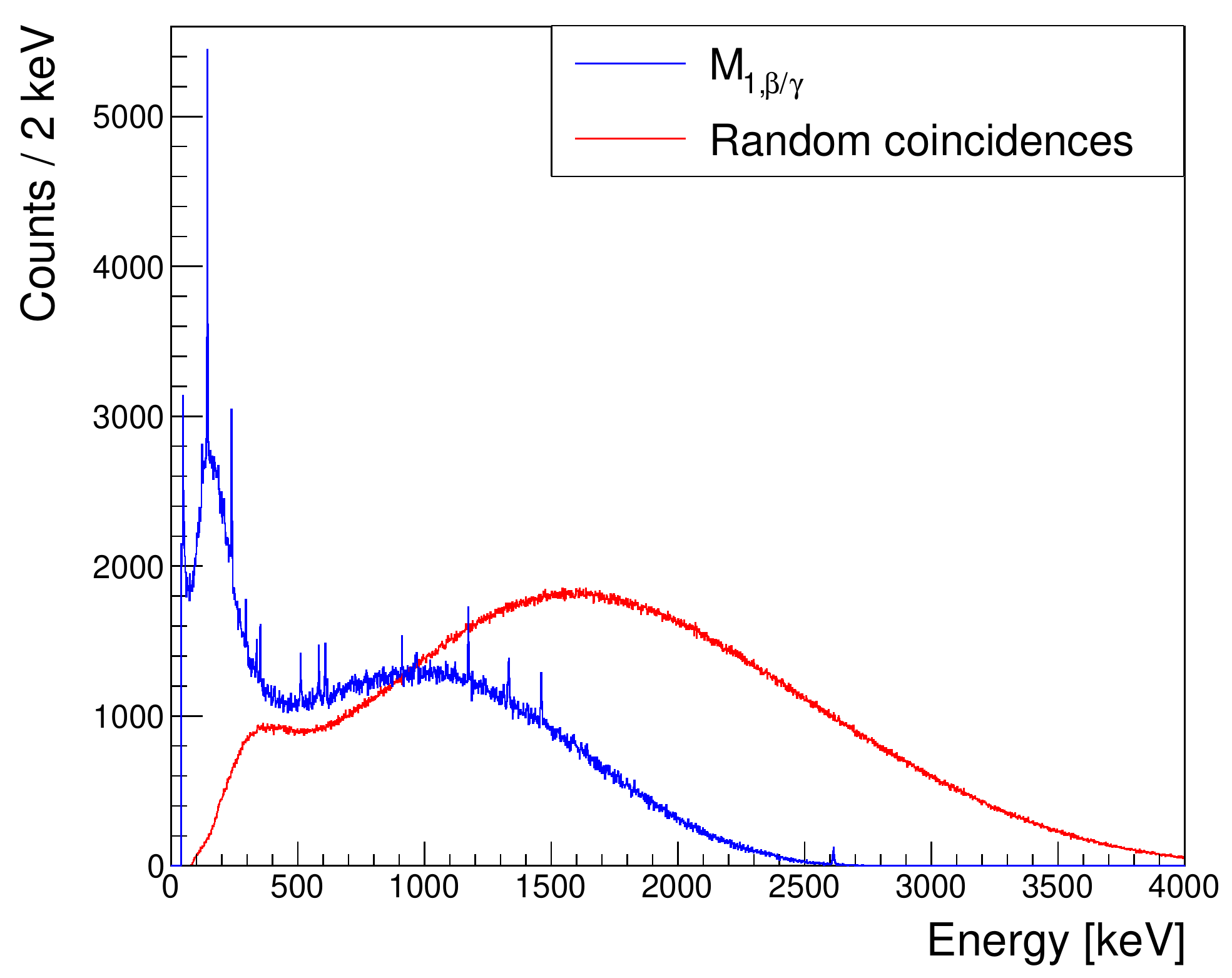}
    \caption{Experimental \M$_{1,\beta/\gamma}$  and random coincidences (obtained by convolution of \M$_{1,\beta/\gamma}$, arbitrary normalization) spectral shapes. We observe  that the  random coincidences distribution is shifted to higher energies (as expected \cite{Chernyak2012,Chernyak:2014ska}) and could cause a background at the ROI.}
    \label{fig:pileup}
\end{figure}

We predict the spectral shape of these events by convolving the experimental  \M$_{1,\beta/\gamma}$ spectrum with itself, i.e. by selecting randomly two energies in the experimental  \M$_{1,\beta/\gamma}$ spectrum and summing them. The  \M$_{1,\beta/\gamma}$  and the resulting random coincidences spectra are shown in Fig. \ref{fig:pileup}.
\\ \indent The expected number of accidentals is then given by:
\begin{equation}
\label{Nrand}
    \hat{N}_{\text{acc}}=N^2 \frac{\Delta t}{t}\times \frac{N_{\text{LMO}}-1}{N_{\text{LMO}}},
\end{equation}
where $N$ is the total number of $\mathcal{M}_1$ events, $\Delta t/t$ is the ratio of the width of the coincidence time window, $\Delta t$ =  $\pm 10 \ \mathrm{ms}$, to the total measurement time and  $N_{\text{LMO}}$ is the number of LMO detectors. 
For the accidental random coincidences we place a prior  as a Gaussian function with mean  $\hat{N}_\text{acc}$ and $\sigma$ $\sqrt{\hat{N}_{\text{acc}}}$. We have used $N$  = 1.2 $\times$ 10$^6$, for a total measurement time of 2.2 $\times$ 10$^7$ s. We include this contribution only in the $\mathcal{M}_2$ spectrum.
 \\\indent The rate of pile-up events is generally lower than the rate of accidentals, but it is also less well constrained as the coincidence time, or effective time resolution, $\Delta t_{\text{eff}}$, is unknown a-priori and determined by the effectiveness of the pulse shape cuts used in the analysis. In general this time resolution will also be dependent on the energy of both the primary and secondary pulse as well as on their separation. However, since we are only interested in events in a narrow range of a high energy region ($\sim$ 3 MeV) we can treat this to a good approximation, as energy independent and simplify Eq. \ref{Nrand} to:
\begin{equation}
\label{Npileup}
    \hat{N}_{\text{pileup}}=N^2 \frac{\Delta t_{\text{eff}}}{t}.
\end{equation}
 For thermal detectors typically the timing resolution is between the inverse sampling frequency and detector rise time. In CUPID-Mo the inverse of the sampling rate is 2~ms and the median value of the rise-time is 24~ms, with 8~ms spread \cite{Armengaud:2019}. Similar LMO crystals to CUPID-Mo, tested at LNGS have achieved $1-2$ ms effective timing resolution \cite{pileup_CUPID} also using PSD only on the LMO channel (as we have done in CUPID-Mo). However, this test has pulses with a higher sampling frequency with respect to CUPID-Mo, different operation temperature and noise conditions and slightly different rise time, so cannot be directly extrapolated.
\\ \indent For the prior on pile-up events in our fit we use a measurement with Th/U calibration sources. We consider all events between 4 -- 5.5 MeV, and from  Eq. \ref{Npileup} 
we can obtain a value for $\Delta t_{\text{eff}}$.  As there are no events in the selected region, we obtain a prior for  $\Delta t_{\text{eff}}$ $<$  7 ms, at 90\% c.i. As zero events are obtained, the corresponding probability density function is an exponential,  that  we  use to place a prior on the pile-up rate. We include this contribution only in the $\mathcal{M}_{1,\beta/\gamma}$ spectrum.

\subsection{Binning and choice of energy intervals}
The energy range of the fit is $100$ -- $4000$ keV for \M$_{1,\beta/\gamma}$,  $400$~--~$4000$ keV for \M$_2$ and  $3000$ -- $10000$ keV for \M$_{1,\alpha}$. We use a variable binning for the three spectra to have enough counts in each bin to minimize the effect of statistical fluctuations. We choose a minimum bin size of 15 keV for \M$_{1,\beta/\gamma}$ and \M$_2$, and  20 keV for  \M$_{1,\alpha}$. We set the minimum number of counts in each bin to be 50 for \M$_{1,\beta/\gamma}$ and 30 for \M$_2$ and  \M$_{1,\alpha}$.  We choose each peak to be fully contained in one bin, to minimize the systematic effect of the detector response on our results. 

\section{Results}
\label{section:results}
\begin{figure*}
    \centering
    \includegraphics[width=1\textwidth]{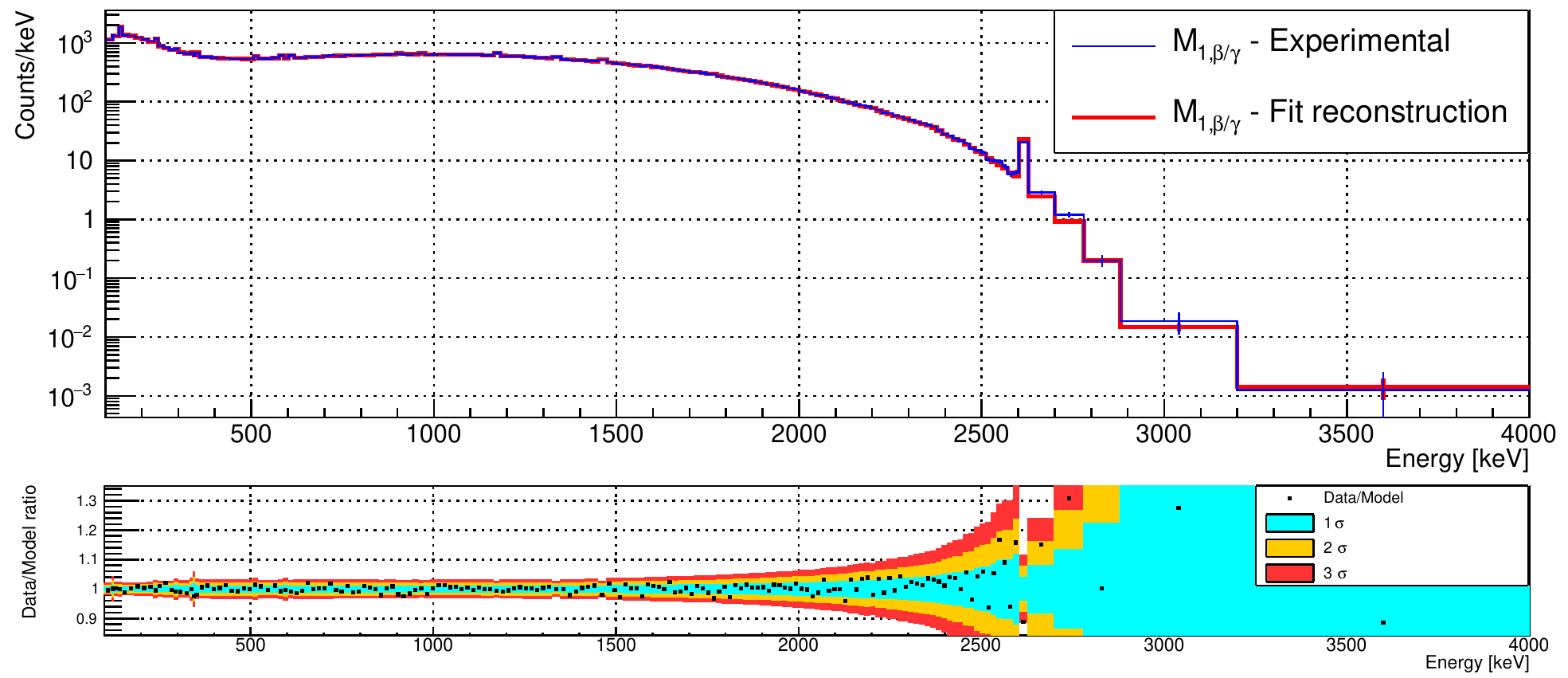}
    \includegraphics[width=1\textwidth]{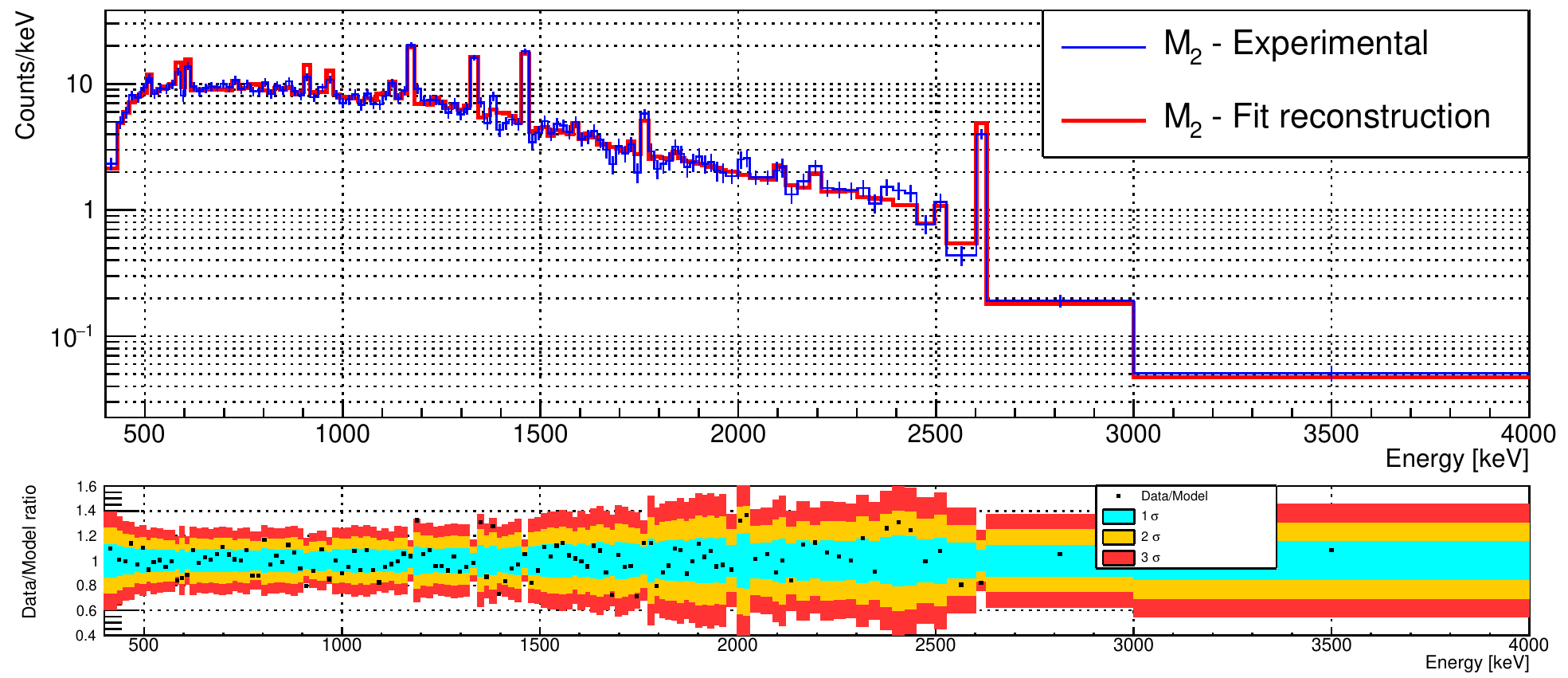}
    \includegraphics[width=1\textwidth]{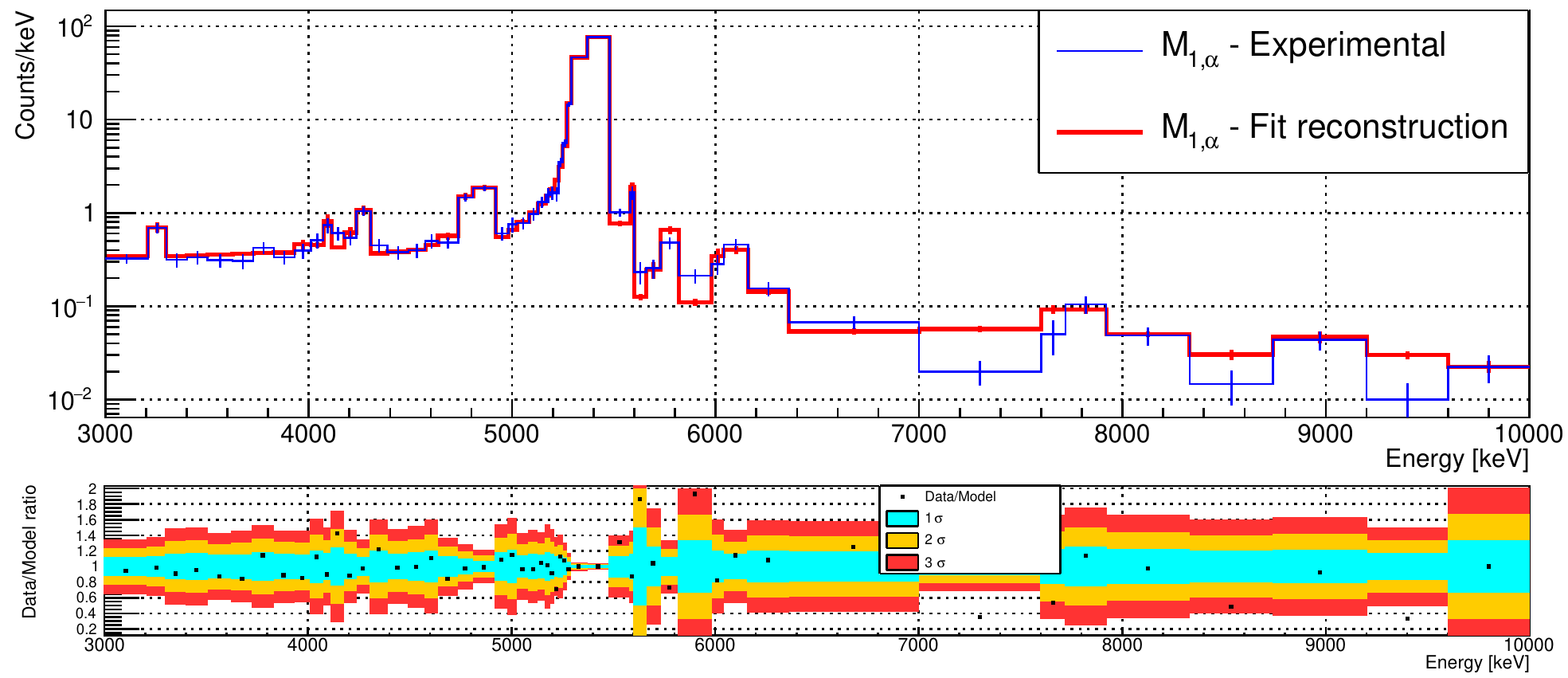}
    \caption{Experimental data and background model simultaneous fit reconstruction of the 3 CUPID-Mo data spectra.  Two upper panels: \M$_{1,\beta/\gamma}$ , $\beta$/$\gamma$'s spectrum with energy deposits in only one detector. Middle panels: \M$_2$, multiplicity 2 events, histogram of the 2 summed energies. Two bottom panels: \M$_{1,\alpha}$, multiplicity 1 events in $\alpha$ energy region. For each one, the lower panel shows the ratio between experimental counts and reconstruction counts for each bin. The colors indicate the uncertainties at $\pm$1, $\pm$2, and $\pm$3 $\sigma$. 
    }
    \label{fig:fit_all}
\end{figure*}

The result of the simultaneous fit of \M$_{1,\beta/\gamma}$, \M$_2$, and \M$_{1,\alpha}$ to the CUPID-Mo data with 2.71 kg $\times$ yr exposure is shown in Fig. \ref{fig:fit_all}. 
Tables \ref{crystal} and \ref{tab:Fit_result} show the fit results, discussed in section \ref{subsection:results_contaminations}. We find that our background model is able to reconstruct well the 3 data spectra. On each spectrum the data over model ratio is shown, where the colors correspond respectively, to $\pm$1, $\pm$2, and $\pm$3 $\sigma$ with:
\begin{equation}
     \sigma_{i,b}=  \frac{\sigma_{\text{data,i,b}}}{n_{\text{model,i,b}}},
\end{equation}
where $n_{\text{model,i,b}}$ is the predicted number of counts in bin $b$ and spectrum $i$ and $\sigma_{\text{data},i,b}$ is the standard deviation of the data in this bin, $\sqrt{n_{\text{data},i,b}}$. 

\begin{figure*}
    \centering
    \includegraphics[width=0.3\textwidth]{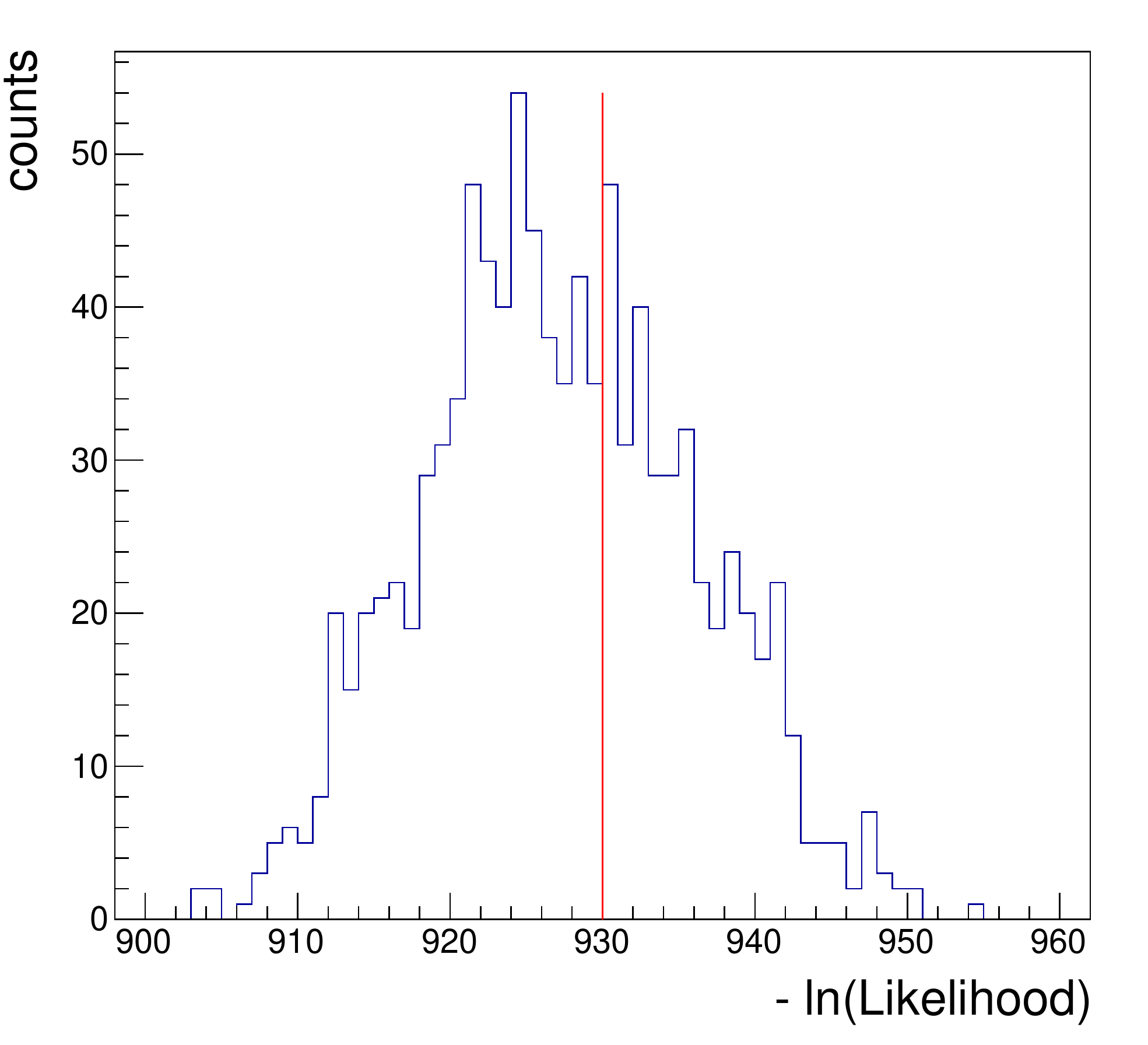}
    \includegraphics[width=0.3\textwidth]{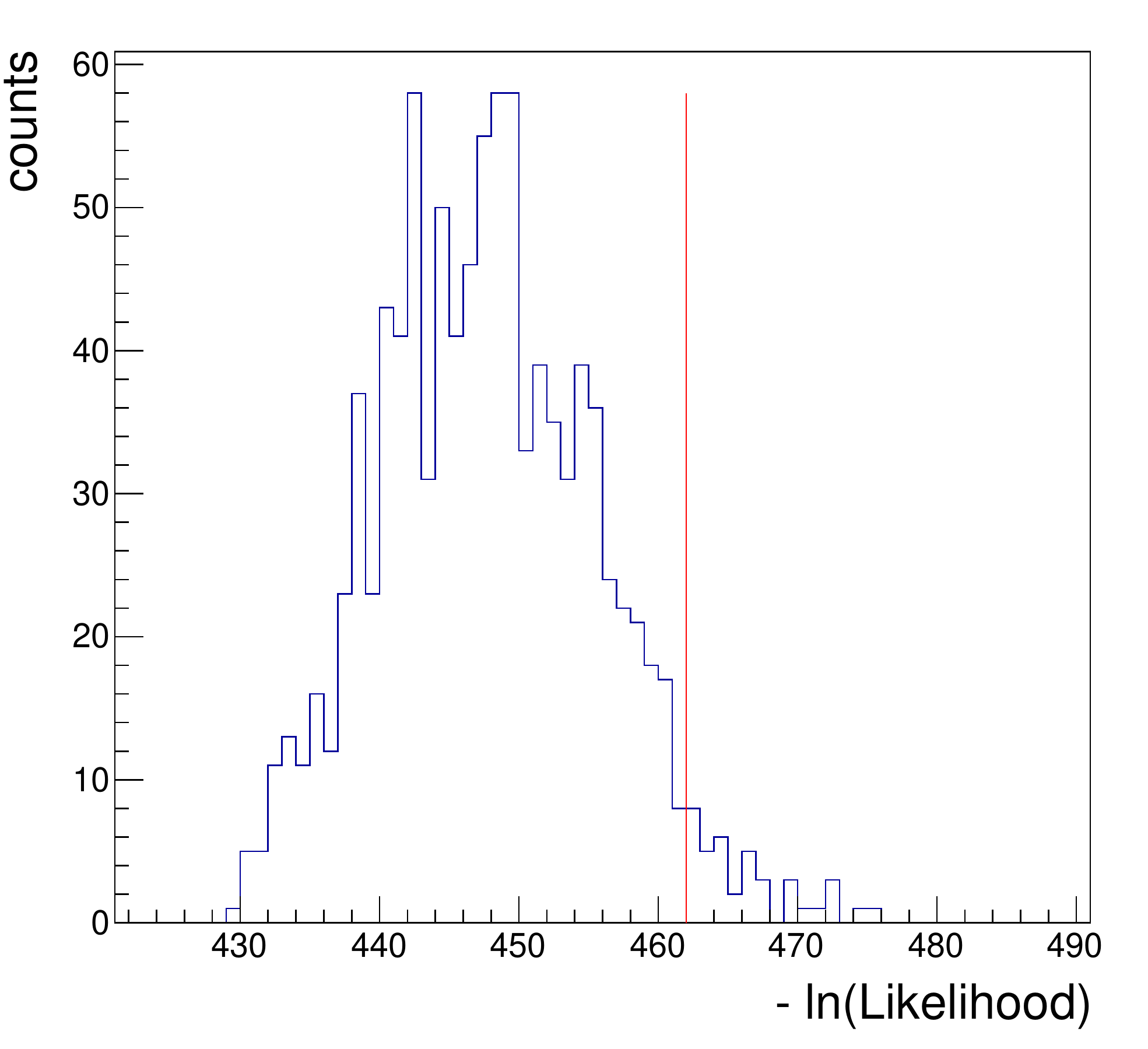}
    \includegraphics[width=0.3\textwidth]{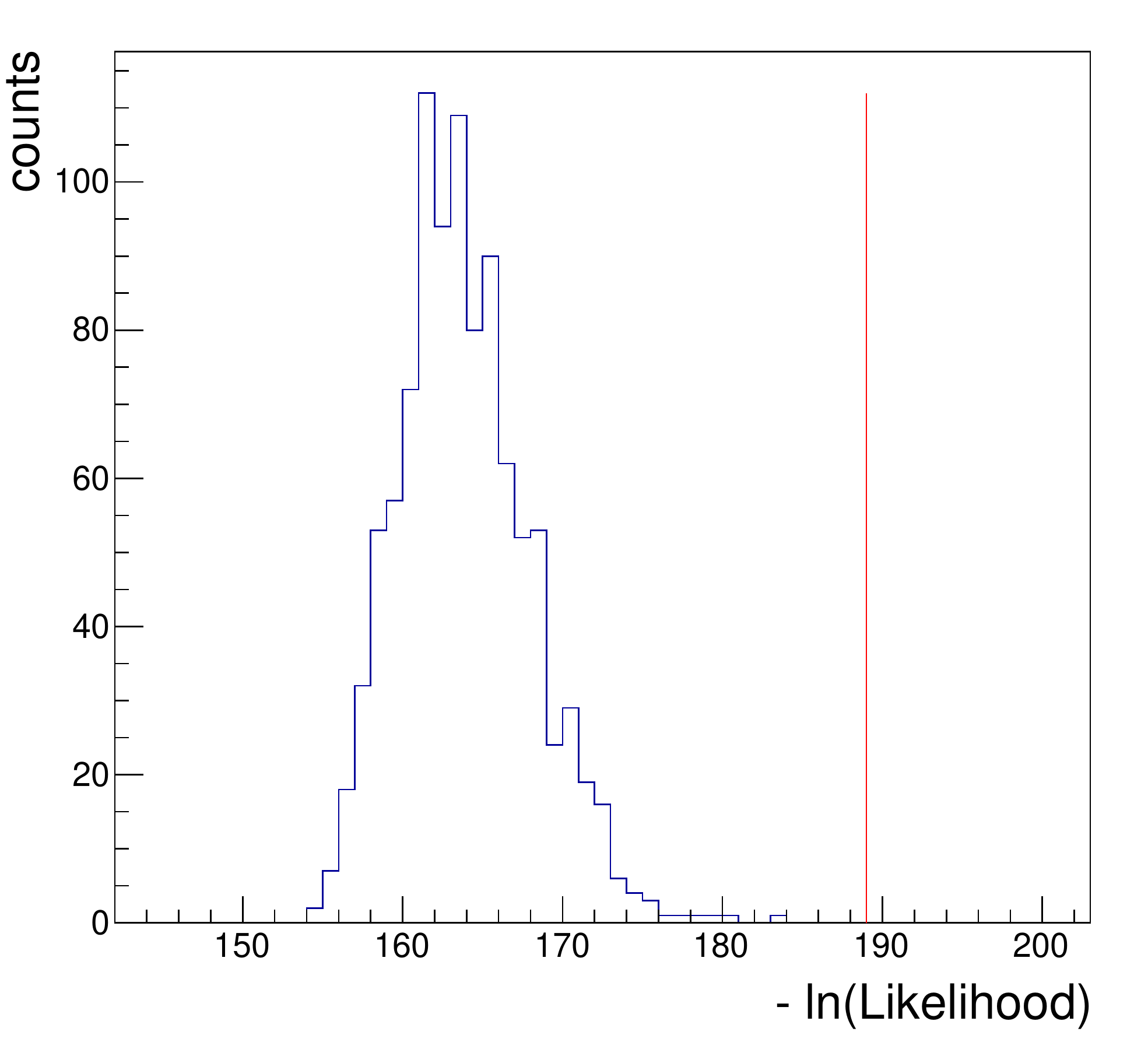}
    \caption{Distribution of $ - \ln \mathcal{L}(\mathcal D|\vec{N})$ from the toys for the \M$_{1,\beta/\gamma}$ (left), \M$_2$ (middle) and  \M$_{1,\alpha}$ (right) spectra. The red line shows the $ - \ln \mathcal{L}(\mathcal D|\vec{N})$ of the reference fit for each of the spectra.}
    \label{fig:toys_MC_chi2}
\end{figure*}

To investigate the goodness of the fit we generate pseudo-experiments, or toy Monte Carlo simulations. We sample randomly according to a Poisson distribution the events in each energy bin of the background model best fit reproduction. We fit independently each pseudo experiment and obtain the likehood $\mathcal{L}(\mathcal D|\vec{N})$, i.e. the probability of the experimental data $\mathcal D$ given the set of parameters $\vec{N}$ of our model. 
We show in Fig. \ref{fig:toys_MC_chi2} the result for  \M$_{1,\beta/\gamma}$, \M$_2$ and  \M$_{1,\alpha}$. The mean of the distributions of 
 \M$_{1,\beta/\gamma}$ and \M$_2$ agree well with the value of the  data. For \M$_{1,\alpha}$ the result demonstrate a modest incompatibility between the data and the model probably arising from an incomplete modelling of $\alpha$ detector response or an $\alpha$ miscalibration. This effect is visible for E>6 MeV in Fig.  \ref{fig:fit_all} bottom panels. This modest incompatibility has driven the choice of a systematic in our model and we have thus performed a fit 
with an energy range  3000 -- 6360 keV for  \M$_{1,\alpha}$. We detail  in subsection \ref{subsection:systematcis} the results. The {\it p} value obtained are $p=0.38$, $p= 0.04$ and  $p \sim$ 0 for \M$_{1,\beta/\gamma}$, \M$_2$ and  \M$_{1,\alpha}$ respectively.


 \subsection{SSD and HSD $2\nu\beta\beta$ decay mechanisms}
The  transition between the ground states of $^{100}$Mo and $^{100}$Ru, with spin parity 0$^{+}$, is realized via virtual $\beta$ transitions through  1$^{+}$ states of the intermediate nucleus $^{100}$Tc. Nuclear theory does not predict a-priori whether this transition is realized dominantly through the 1$^{+}$ ground state (SSD hypothesis) or through higher excited states of $^{100}$Tc (HSD hypothesis) \cite{NEMO3_Mo_detailed_studies}.
\\ \indent We have found that the SSD mechanism of $2\nu\beta\beta$ decay to $^{100}$Ru ground state reproduces fairly well the data 
with a $p=0.38$, while the HSD model does not,  $p \sim 0$.
Since our data clearly favours SSD over HSD mechanism for $2\nu\beta\beta$, we have used the SSD model in our final fit.

\subsection{Contaminations derived from the fit}
\label{subsection:results_contaminations}
\subsubsection{LMO crystal contaminations}
\begin{figure*}
    \centering
    \includegraphics[width=\textwidth]{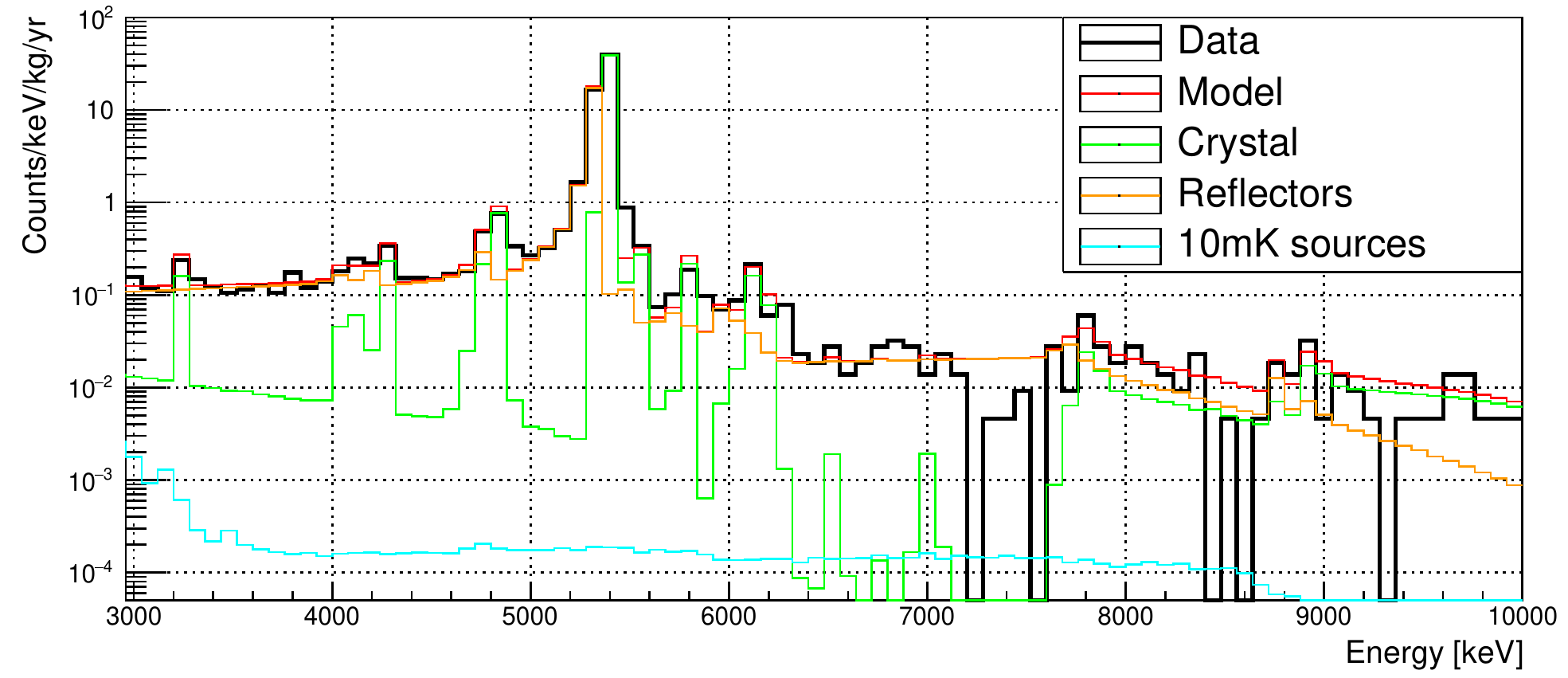}
    \caption{Experimental 
    \M$_{1,\alpha}$ spectrum reconstruction showing the components of the $\mathcal{M}_{1,\alpha}$ background model fit. Crystal and Reflector contaminations include bulk and surface. The surface contaminations are modelled with an exponential density profile and $\lambda$ = 10 nm parameter depth. The peaks in the spectrum are described by the radioimpurities in the crystal and the continuum by the ones in the bulk of the Reflectors. The small contribution from 10 mK sources corresponds to the holders.}
    \label{fig:fit_components_alpha}
\end{figure*}
 
The \M$_{1,\alpha}$ spectrum is populated by $\alpha$ decays occurring in the crystals and in the elements directly facing them. As described in section \ref{subsection:peaks} we included  bulk and surface contaminations in the crystals in our fit. We show in Fig. \ref{fig:fit_components_alpha}, the resulting components. Since we do not observe clear $\alpha$-energy (NR escape) peaks due to the very low levels of contaminations and thus limited statistics, bulk and surface contaminations are anticorrelated. We performed studies concerning the effect of the location of the contaminations on bulk or surface in the fit results, which we discuss later. 
 \\ \indent The largest peak in the $\alpha$ region is the $^{210}$Po peak. This peak is largely described by the $Q$-value component of the crystal bulk. 
 For the $^{210}$Po in order to fit as much as possible the particular shape peak in addition to 10 nm, implantation depths of 1 $\mu$m and 1 nm are used in the crystals, and implantation depths of 100 nm and 1 $\mu$m are used in the Reflectors.
In Fig. \ref{fig:fit_components_alpha} the left tail of the $^{210}$Po peak is described by the surface  contamination on the Reflectors. 
 \\ \indent The summary of the crystal activities extracted from the fit is presented in Table \ref{crystal}. The LMO crystal contaminations by radionuclides from the $^{238}$U and $^{232}$Th chains are all below 1 $\mu$Bq/kg. As a study of the effect of the bulk versus surface location, we performed a fit without surface contaminations. These results show that, even under this extreme assumption, the results on the bulk activities do not vary significantly.  As shown in Fig. \ref{fig:data} the peak at 4.8 MeV contains $^{234}$U and $^{226}$Ra alpha decays.  In this analysis this peak is ascribed to $^{234}$U, with a significant uncertainty (reported in Table \ref{crystal}) in the resulting contamination due to the anticorrelation with the  $^{226}$Ra contribution. Additionally, in this peak we could have a contribution from the neutron capture in $^{6}$Li \cite{Armengaud:2017}. Neutrons captured in $^{6}$Li produce an alpha particle plus tritium, $^{6}$Li(n,$\alpha$)$^{3}$H, with a total energy 4.782 MeV. We note also that the level of $^{228}$Ra is not constrained by any $\alpha$ peak.
  \\ \indent There is clearly a larger $^{210}$Po contribution than the rest of the $^{238}$U chain, at the level of  96~$\mu$Bq/kg, possibly introduced during the purification of the enriched material \cite{Armengaud:2015}. There are also traces of $^{190}$Pt, caused by the crystal growth in a platinum crucible \cite{grigor} and we find $^{40}$K and  $^{90}$Sr+$^{90}$Y at the level of some hundreds of $\mu$Bq/kg.
We note that  $^{210}$Pb, $^{87}$Rb, $^{90}$Sr+$^{90}$Y and $^{40}$K do not represent a potential background for \onbb search, as the  $Q{_{\beta}}$ of these radioisotopes is much lower than the \onbb ROI at 3 MeV.
\\ \indent We show in Table \ref{crystal} (bottom) the surface contaminations of the crystals derived from the fit. We studied the effect of including also a contribution with a depth parameter of $10 \mu$m (i.e., including surface contaminations with $\lambda$ = 10 nm and 10 $\mu$m) and the decay activitiy is shown in the third column of the table. The results are compatible with the fit including only 10 nm contributions. 
We observe clear anti-correlation for a given decay chain between the bulk and the surface contaminants in the crystal, but also with the surface of  the Reflectors. These anti-correlations are taken into account in the uncertainties given in Table \ref{crystal}.

\begin{table}
\small
\caption{Radioactive contaminations of the LMO crystals derived from the background model of the CUPID-Mo data, with 2.71 kg $\times$ yr exposure. The upper table shows the bulk activites. We report also the results under the assumption of no surface contaminations, to study the effect in the fit of the anticorrelation between bulk and surface activities. The lower table shows the surface activities, we give the activities with MC simulation with 10 nm implantation depth (see text). The effect of including a contribution with a depth parameter of 10 $\mu$m  is shown on the last column.}
\label{crystal}       
\renewcommand{\arraystretch}{1.4}
\begin{tabular}{llll}
\hline\noalign{\smallskip}
  Chain       & Nuclide   & Bulk Activity             &  w/o surface cont.
  \\
              &            &     [$\mu$Bq/kg]    &  [$\mu$Bq/kg] 
              \\
\hline

\hline
   $^{232}$Th &  $^{232}$Th   & $< 0.22$ & $0.18^{+0.09}_{-0.05}$ 
   \\ 
             & $^{228}$Ra to $^{228}$Th & $< 79$ & $< 98$    \\
             & $^{228}$Th to $^{208}$Pb & $0.43^{+0.16}_{-0.15}$ & $0.57 \pm 0.07$ 
             \\
\hline
  $^{238}$U & $^{238}$U to $^{234}$U   & $0.41^{+0.16}_{-0.28}$ & $0.59^{+0.12}_{-0.11}$
  \\
            & $^{234}$U           & $1.15^{+0.33}_{-0.70}$ & $1.59 \pm 0.20$ 
            \\
            &  $^{230}$Th  & $<0.58$ & $0.47^{+0.23}_{-0.24}$  
            \\
           & $^{226}$Ra to $^{210}$Pb &  $<0.21$    & $0.39 \pm 0.06 $         
           \\
          &  $^{210}$Pb  & $96^{+6}_{-27}$  &     $105 \pm 1$ 
          \\
            \hline
         &  $^{190}$Pt  & $0.39^{+0.11}_{-0.10}$  & \\
        
          &  $^{87}$Rb  & $< 103$   & 
          \\
           & $^{90}$Sr-$^{90}$Y  & $159^{+38}_{-34}$ & 
           \\
           & $^{40}$K  & $41^{+29}_{-22}$ 
           &      \\
           \hline\hline
     &     &  \multicolumn{2}{c}{Surface Activity [nBq/cm$^2$]} \\
     &  & 10 nm     &  10 nm + 10 $\mu$m       \\
\noalign{\smallskip}\hline\noalign{\smallskip}
&$^{232}$Th  & $<1.3$& $< 1.1$ \\
&$^{228}$Ra to $^{228}$Th & $< 389$ & $<449$\\
&$^{228}$Th to $^{208}$Pb & $<2.5$ & $0.9^{+0.9}_{-0.6} $ \\           
 \hline          
    & $^{238}$U to $^{234}$U & $<2.9$ & $<$ 2.4  \\
      & $^{234}$U  & $<7.3$ &  $< 5.9$ \\ 
      &$^{230}$Th  & $<2.2$ & $<2.3$ \\
&$^{226}$Ra to $^{210}$Pb & $2.0 \pm 0.5$ & $<2.1$ \\
&$^{210}$Pb to $^{206}$Pb$^a$ & $62^{+109}_{-31}$  \\
 \hline          
           
  \multicolumn{4}{l}{$^{a}$ includes 1 $\mu$m and 1 nm implantation depth (see text for details)}
\end{tabular}
\end{table}

\subsubsection{Radioactive contaminations of the setup components} 
\begin{figure*}
    \centering
    \includegraphics[width=\textwidth]{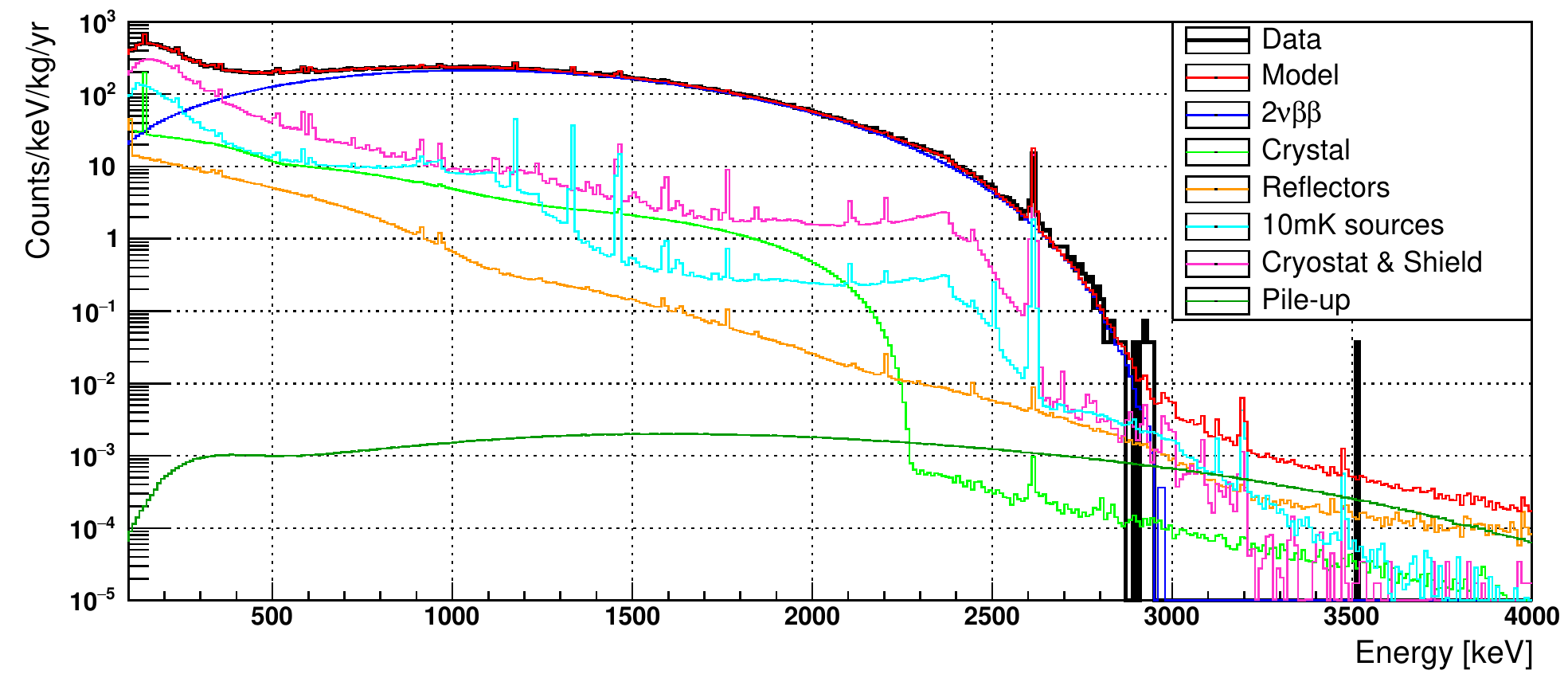}
    \caption{Background sources reconstructing the experimental  $\mathcal{M}_{1,\beta/\gamma}$ spectrum, grouped by source location. In blue, \nnbb is the dominant contribution in [350--3000] keV. The most important contribution from the materials, below 3 MeV, are the cryostat and shields, shown in magenta.}
    \label{fig:fit_components_M1}
\end{figure*}

A list of sources included in the fit and their resulting activities obtained from the marginalised mode and 68\% c.i. are shown in Table \ref{tab:Fit_result}.
\\ \indent The derived activities for the component called {\it{Reflectors}} are mainly constrained by the fit of the continuum in the $\alpha$ region. The values are larger than the measured radioactivities of the reflectors themselves, in particular, in $^{226}$Ra.  We remind that this component takes into account all elements directly facing the crystals: PTFE, NTDs, LDs, bonding wires. A contamination of the reflecting foils introduced during the detector assembly could be conceived, explaining the activities obtained in the fit.
\\ \indent Concerning  the surface activity on the reflecting foils, we performed a measurement with the BiPo-3 detector \cite{BiPo} which measures $^{214}$Bi and $^{208}$Tl levels through delay coincidences in the Bi-Po cascades. We can also convert the ICPMS results of the bulk measurement by assigning all the contamination to the surface. 
The surface activity of the {\it{Reflectors}} derived from the fit agrees well within uncertainties with both measurements.
\\ \indent The derived activities in the {\it{Kapton cables}}, the {\it{Connectors}}, the {\it{Brass Screws}} and the {\it{Copper supports}} agree  well with the measured values.
For the {\it{Cryostat Screens}} the activities obtained in the fit are higher than the measured levels from the raw copper. This points out to an additional contamination introduced during the fabrication of the screens for example due to the weldings.  In particular, we have identified from the experimental data that the detectors facing the weldings in the cryostat screens have higher rates in the 2615~keV peak of $^{208}$Tl.  \\
\indent The {\it{Screen 300 K}} accounts for the residual environmental $\gamma$'s and the radon present in the gap between the outermost cryostat screen and the external lead shielding. The $^{226}$Ra contamination derived from the fit shown in Table  \ref{tab:Fit_result} can be translated into a radon level concentration resulting in (22~$\pm$~3)~mBq/m${^3}$, which is  in good agreement with measurements of 20 mBq/m${^3}$ provided by the radon mitigation system in the LSM \cite{Hodak}. 
\\ \indent
Figure \ref{fig:fit_components_M1} shows the breakdown of the components in the fit of \M$_{1,\beta/\gamma}$. In the region $0.8-3$ MeV the dominant contribution is the \nnbb from $^{100}$Mo and the most important contributions from the radioactivity in the materials are the cryostat and shields. We discuss in the next section the main sources in the \onbb region.

\begin{table*}
\small
\caption{Radioactive contaminations of the setup components derived from the posterior distribution of the background model fit. Uniform, non-informative priors are used except for the $^{228}$Th, $^{226}$Ra and $^{40}$K contaminants in the springs. 
For surface contaminations, the simulated depth is 10 nm. The last column shows the activities from screening measurements when available (see Tables \ref{cupid-mo-screening} and \ref{edw-screening} in section \ref{section:background_sources}). }
\label{tab:Fit_result}       
\renewcommand{\arraystretch}{1.4}
\begin{minipage}{20cm}
\begin{tabular}{c|c|c|c}
Component                              & Bulk             & Posterior   &  Activity from screening     \\
                                       &                   &  [mBq/kg]   & [mBq/kg]                      \\ 
\hline
\hline
Reflectors$^a$ &  $^{238}$U to $^{210}$Pb  & $9.2 \pm 1.0$   & Refl. only: $0.17 \pm 0.05$    \\ 
               &  $^{210}$Pb               & $<17$           &                                            \\
               &  $^{232}$Th to $^{208}$Pb & $<2.3$          &   Refl. only:  $0.05 \pm 0.01$              \\
\hline
Springs        &  $^{228}$Ac               & $<217$          &         \\
               &  $^{228}$Th to $^{208}$Pb & $20 \pm 5$  &      $21\pm 5$        \\
               &  $^{226}$Ra to $^{210}$Pb & $10 \pm 3$      &      $11\pm 3$         \\
               &  $^{40}$K                 & $3440^{+450}_{-340}$ &  $3600 \pm 400$     \\
\hline
Kapton cables &   $^{228}$Ac               & $<139$          &      \\ 
              &   $^{228}$Th to $^{208}$Pb & $<28$           &      $15\pm 10$            \\
              &   $^{226}$Ra to $^{210}$Pb & $<13$           &      $8\pm 6$      \\
\hline
Connectors$^b$ &  $^{228}$Ac               & $<442$          &                    \\
              &   $^{228}$Th to $^{208}$Pb & $<339$          &      $82\pm 38$   \\
              &   $^{226}$Ra to $^{210}$Pb & $<169$          &      $15 \pm 8$ \\ 
\hline
 Brass Screws &   $^{228}$Ac               & $<24$           &         \\
              &   $^{228}$Th to $^{208}$Pb & $<18$           &      $3.5\pm 0.9$    \\
              &   $^{210}$Bi$^c$           & $(3.0 \pm 0.3) \times 10^4$ &  $620 \pm 254$     \\
\hline
Copper supports & $^{228}$Ac               & $<0.051$        &     \\ 
              &  $^{228}$Th to $^{208}$Pb  & $<0.052$        &     $0.024 \pm 0.012$ \\ 
              &  $^{226}$Ra to $^{210}$Pb  & $<0.019$        &      $<0.04$     \\
              &  $^{60}$Co$^d$             & $0.47 \pm 0.02$ &   $0.04$         \\
              &  $^{57}$Co$^d$             & $0.029  \pm 0.005$     &            \\
\hline
Cryostat Screens & $^{228}$Ac              &$<0.38$          &           \\
              &  $^{228}$Th to $^{208}$Pb  &$< 0.40$         &  $0.024 \pm 0.012$        \\
              &  $^{226}$Ra  to $^{210}$Pb &$<0.15$          &  $< 0.04$         \\
\hline
PE 1K$^e$     & $^{228}$Ac                 &$<4.4$           &$0.5 \pm 0.2$        \\
              & $^{228}$Th to $^{208}$Pb   &$2.2^{+2.1}_{-1.6}$&   $0.3 \pm 0.1$            \\
              & $^{226}$Ra to $^{210}$Pb   &$<2.1$           & $0.65 \pm 0.08$               \\
\hline
Screen 300K   & $^{228}$Ac to $^{208}$Pb   &$(203^{+48}_{-51}$) mBq &                   \\
              & $^{226}$Ra to $^{210}$Pb   & $(94 \pm 13)$ mBq &   
              \\
              & $^{40}$K                   & $(3200 \pm 400)$ mBq      &           \\
\hline\hline
 Component     & Surface                    &  Posterior    & Activity from screening    \\
               &                            &  [nBq/cm$^{2}$]    & [nBq/cm$^{2}$]     \\ 
                                        \hline
Reflectors$^a$ &  $^{238}$U to $^{234}$U    &  $2.7_{-1.6}^{+1.9}$             &     \\
               &  $^{234}$U                 &  $< 9.5$                         &     \\
               &  $^{230}$Th                &  $<3.5$                          &     \\
               &  $^{226}$Ra to $^{210}$Pb  &  $3.4_{-1.2}^{+1.5}$             &  $(1.0 \pm 0.4)^f$ / $(1.7 \pm 0.5)^g$   \\
               &  $^{210}$Pb to $^{206}$Pb  &  $1034_{-33}^{+26}$               &     \\
               &  $^{232}$Th                &  $<3.9$                          &     \\
               &  $^{228}$Ra to $^{228}$Th  &  $<504$                          &     \\
               &  $^{228}$Th to $^{208}$Pb  &  $2.6_{-1.5}^{+1.4}$             &  $(1.1 \pm 0.4)^g$    \\ \hline
\end{tabular}
\end{minipage}
$^a$ Reflectors take into account all passive elements directly facing the crystals: Reflecting foils, PTFE, bonding wires, heaters; \\
$^b$ Connectors refer to MillMax connectors plus Kapton connectors;\\
$^c$ The $^{210}$Bi in the Brass Screws accounts for this contamination in all 10 mK (Cables, Connectors, Springs, Copper supports) and infrastructure sources (cryostat screens and PE 1K); \\
$^d$ Co in Copper supports account for this contamination also in Cryostat screens;\\
$^e$ 1K PE accounts for all sources below the 10 mK stage, e.g., 300 K electronics, dilution unit;\\
$^f$ $^{214}$Bi surface measurement with the BiPo-3 detector;\\
$^g$ Extrapolation from ICPMS measurement, assuming all contamination on surface. 
\end{table*}

\subsection{Background index in the $^{100}$Mo $0\nu\beta\beta$ ROI}
\label{subsection:results_ROI}
\begin{figure}[h!]
\centering
\includegraphics[width=0.45\textwidth]{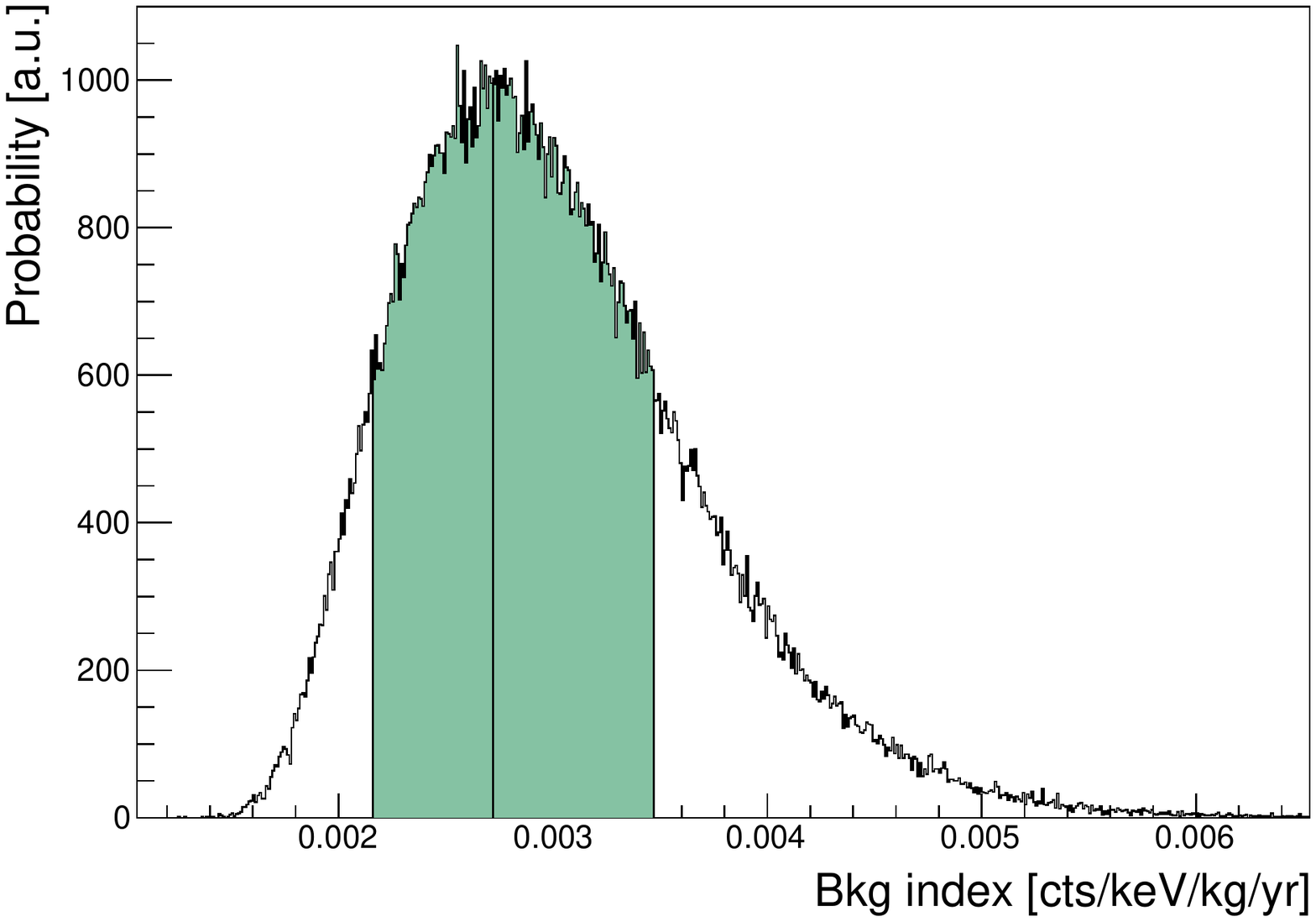}
\caption{Posterior distribution of background index, showing the mode and the smallest 68.3\% c.i., $2.7_{-0.6}^{+0.7}\times 10^{-3}$ counts/keV/kg/yr.}
\label{bkg-index}
\end{figure}

\begin{figure*}
    \centering
    \includegraphics[width=1\textwidth]{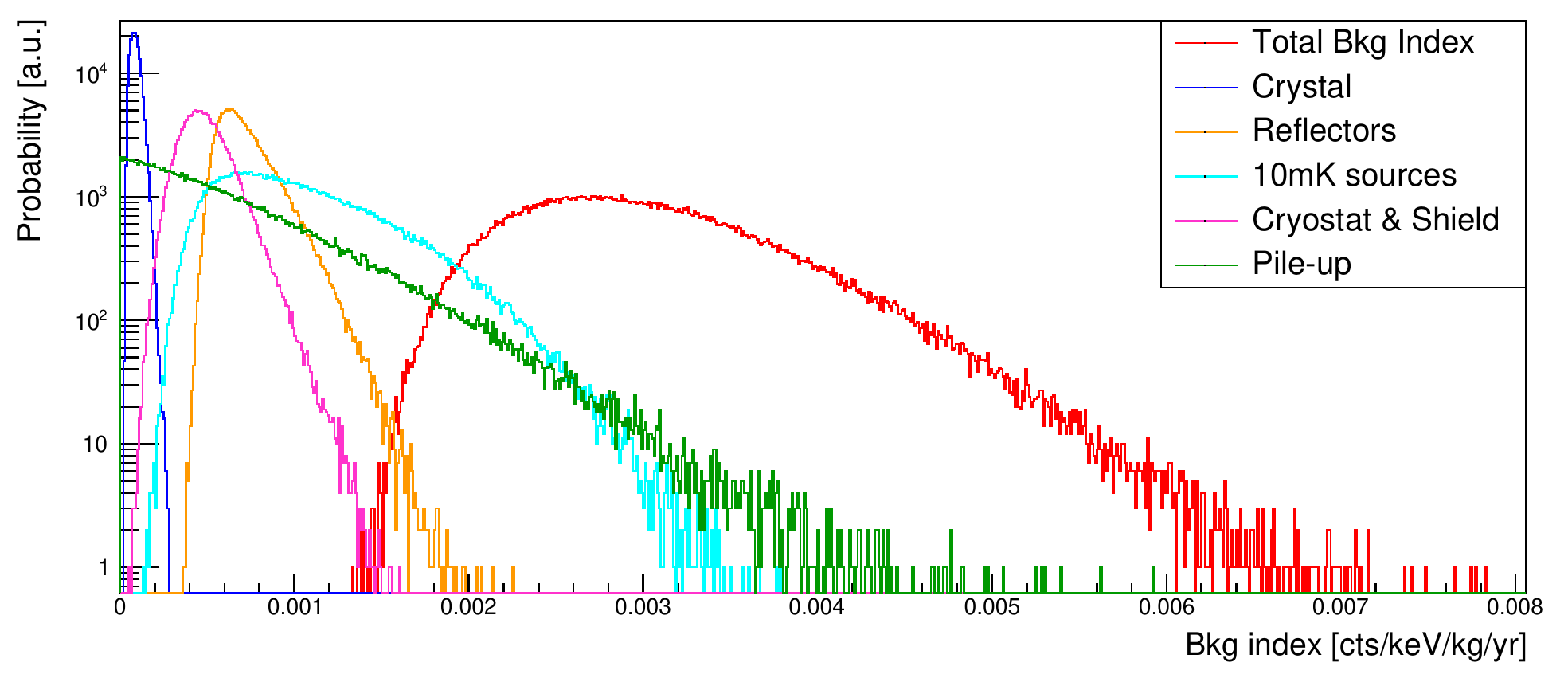}
    \caption{Posterior distributions of background index of the several background sources grouped by source location. Also shown is the full posterior distribution. }
    \label{Breakdown}
\end{figure*}

\begin{figure*}
\centering
\includegraphics[width=1\textwidth]{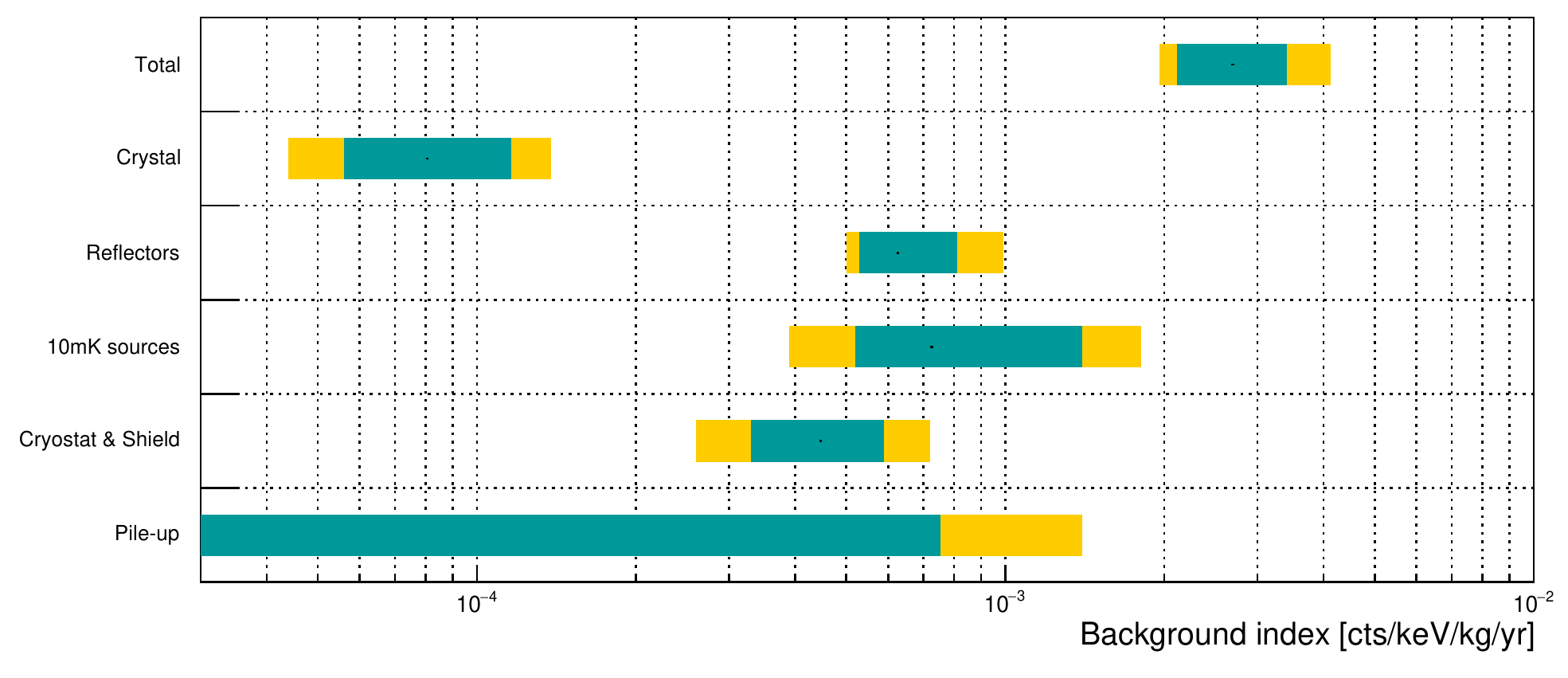}
\caption{Background index for the various groups of sources. The values are extracted from the mode of each distribution of Fig. \ref{Breakdown}, with their respective uncertainties.
The green bars correspond to the smallest 68.3\% interval around the mode,  and the yellow bars to the smallest 90\% interval around the mode. For the pile-up  the distribution is compatible with zero, thus we give an upper limit to 68.3\% c.i. in green and 90\% c.i. in yellow.}
\label{bkg_budget}
\end{figure*}

We use our simultaneous fit to reconstruct the background index in the $0\nu\beta\beta$ region of interest. We chose to calculate the background index in the region $\pm 15$ keV around 3034 keV.
We sample directly the full posterior distribution produced by JAGS for each step $i$ in the Markov Chain by computing: 
\begin{equation}
    b_i =\sum_{\text{j=1}}^{\text{67}} \text{Pois}(N_j)\frac{w_{i,j}}{\Delta E \times Mt }.
\end{equation}
Here $b_i$ is the background index in the $0\nu\beta\beta$ region of interest, $N_j$ is the integral of the spectrum of MC source {\it j} in the ROI, $w_{i,j}$ is the weight for source {\it j} in step {\it i}, $\Delta E$ is the width of the ROI and $Mt$ is the experimental exposure. The sum goes over all background sources. The MC simulations are themselves the result of a stochastic process they have a statistical uncertainty, this is accounted for by Poisson smearing the MC ROI integrals per step of the Markov Chain. We then use the distribution of $b_i$ for the full Markov Chain to estimate the marginalised posterior distribution of the background index.
We perform this calculation for our maximal model with all parameters, we therefore naturally marginalise over all possible combinations of activities (for example surface or bulk radio-purity) consistent with our experimental data accounting for the systematic uncertainty due to source localisation.
\\ \indent From this calculation we extract the marginalised posterior of the background index shown in Fig. \ref{bkg-index}. This results in a measurement (mode $\pm $ smallest 68\% interval) of:
\begin{equation}
    b = 2.7^{+0.7}_{-0.6} \times 10^{-3} \text{counts/keV/kg/yr}.
    \label{BI}
\end{equation}
or, in terms of the number of moles of isotope, mol$_{\text{iso}}$, and energy resolution, $\Delta E_{\text{FWHM}}$:
\begin{equation}
     \mathcal{B}= 3.7^{+0.9}_{-0.8}~\times 10^{-3}  
      \text{counts/$\Delta E_{\text{FWHM}}$/mol$_{\text{iso}}$/yr}
    \label{BI}
\end{equation}
This is the lowest  background index achieved in a bolometric \onbb decay experiment.
\\
\indent
Next we reconstruct the contributions to the experimental background. We divide sources into five categories:
\begin{itemize}
    \item Crystals U/Th;
    \item Pile-up;
    \item Reflectors;
    \item 10 mK sources;
    \item Cryostat and shields.
\end{itemize}
We emphasise that only the first three sources are relevant to CUPID. In the CUPID baseline the reflective foil is removed to improve background rejection. However, as it was noted before {\it Reflectors} include all the elements directly facing the crystals, PTFE, bonding wires, heaters. These elements will remain in CUPID. The final two are caused by materials in the EDELWEISS cryostat which is optimised for a dark matter rather than $0\nu\beta\beta$ decay search.
The posterior distributions of background index from each source are shown in Fig. \ref{Breakdown}. We derive the background index for each of the sources in the same way as for the full posterior. 
We find that the crystals give the smallest contribution, with a background index:
\begin{equation}
     8.1^{+3.5}_{-2.5} \times 10^{-5} \text{counts/keV/kg/yr}.
\end{equation}
As shown in  Fig. \ref{Breakdown}, the posterior probability for pile-up allows us to set an upper limit for its background index,  $<1.4 \times 10^{-3}$ \text{counts/keV/kg/yr} (90\% c.i.). This is potentially the main background contribution, in particular due to the low CUPID-Mo sampling frequency (500 Hz) and lack of optimised cuts to remove pile-up. In CUPID, heat and light signals will be exploited together with optimised algorithms to remove pile-up events (see for example \cite{fantini_pu}).
Figure \ref{bkg_budget} gives the background index  extracted from Fig. \ref{Breakdown} for each of the grouped components. They are obtained from the mode and the smallest 68.3\% interval.  
For the pile-up the smallest 68.3\% interval is compatible with zero, thus an upper limit is presented. 


\subsection{Systematics}
\label{subsection:systematcis}
To check the stability of the model and the systematic uncertainties, we perform a series of different fits.
To take into account the systematic uncertainty due to MC statistics, we add a nuisance parameter in Eq. \ref{likelihood}:
\begin{align}
\nonumber
   \text{ln}(\mathcal{L}\Big(D|(\vec{N})\Big)) &= \\ \nonumber \sum_{i=1}^{3} \sum_{b=1}^{N_b(i)} &\ln{(\text{Poiss}(n_{i,b};f_i(E_b;\vec{N})))} \\ 
   +&\ln{(\text{Poiss}(N^{MC}_{j,i,b}; \hat{N^{MC}_{j,i,b}}))},
\end{align}
where, $N^{MC}_{j,i,b}$ is the number of MC events in bin $b$ of source $j$ in spectra $i$, and $\hat{N^{MC}_{i,b}}$ is the expected number. These nuisance parameters added to the model  account for the integer Poisson fluctuations in the MC. We find that the fit remains largely unchanged with only a small change in the value of the background index. 
\\ \indent To check the stability of the fit, we perform different fits varying the binning, the energy fit region, the choice of background sources and, in particular, the bulk and surface contaminations in the crystals, as follow:

\begin{itemize}
 \item Binning: we repeat the fit with 1, 2 and 20 keV fixed binning in $\mathcal M_{1,\beta/\gamma}$ and \M$_2$. In all cases, the overall goodness of the fit remains, and the value of the background index is compatible within uncertainties to that of the reference fit, as shown in Table \ref{systematics}. We did not repeat the fit with 1, 2 and 20 keV   on \M$_{1,\alpha}$ due to the low  statistics in each bin of the data;
    \item Fit energy region: our reference fit extends from 100~keV to 4~MeV for \M$_{1,\beta/\gamma}$ spectrum.
We vary the energy threshold to 200 keV  and find that the background index only varies slightly;
    
    \item Choice of  background sources: our calculation of the background index is naturally marginalising over this uncertainty (see section \ref{subsection:results_ROI}). However, as an additional check  we perform the fit without including the crystal bulk contribution for the U and Th chains. The values of the activities of the sources change, but the goodness of the fit remains very similar and the value of the background index remains almost unchanged. We then remove the crystal surface contamination and still obtain a background index compatible within uncertainties to that of the reference fit;

    \item Energy region of \M$_{1,\alpha}$ fit: our reference fit extends from [3000 - 10000] keV. As described at the beginning of section \ref{section:results} the  \M$_{1,\alpha}$ fit shows a modest incompatibility between the data and the model, mainly in the region E>6~MeV. We thus performed a fit in [3000 -- 6360] keV to account for this incompatibility as a systematic uncertainty in our model. In doing so, the U and Th contributions in the crystal get more degenerated, resulting in an increase of the Th contamination assigned in the fit. Still the background index is compatible, within uncertainties, to that of the reference fit.
    \end{itemize}
 \indent The results of these tests are summarized in Table \ref{systematics}. 
 As argued above, the result given in Eq. \ref{BI} is naturally marginalising over the uncertainty on the choice of the background sources.  
 Considering all tests in Table \ref{systematics} as a systematic uncertainty (with 2 keV fixed binning) and adding them in quadrature,  the background index in (3034 $\pm$ 15) keV results in:
   \begin{equation}
    b = 2.7^{+0.7}_{-0.6}\text{(stat)} ^{+1.1}_{-0.5}\text{(syst)} \times 10^{-3} \text{counts/keV/kg/yr}.
\end{equation}
or:
\begin{equation}
  \mathcal{B}= 3.7^{+0.9}_{-0.8}\text{(stat)}^{+1.5}_{-0.7}~\text{(syst)}~\times 10^{-3}
      \text{counts/$\Delta E_{\text{FWHM}}$/mol$_{\text{iso}}$/yr}
    \label{BI}
\end{equation}

\begin{table}
\small
\caption{Background Index in ROI for different fits. The tests allow to check the stability of the model and assess the systematic uncertainties. }
\label{systematics}       
\renewcommand{\arraystretch}{1.4}

\begin{tabular}{ll}
\hline\noalign{\smallskip} 
Fit &  Background Index \\
     & [$10^{-3}$ \text{cts/keV/kg/yr}]\\
\noalign{\smallskip}\hline\noalign{\smallskip}
Reference fit &  $2.7 \substack{+0.7\\ -0.6}$ \\
\noalign{\smallskip}\hline\noalign{\smallskip}
\M$_{1,\beta/\gamma}$ threshold = 200 keV                   &  $2.8_{-0.6}^{+0.7}$ \\
1 keV fixed binning for \M$_{1,\beta /\gamma}$ and \M$_{2}$  &  $2.5_{-0.5}^{+0.6}$ \\
2 keV fixed binning for \M$_{1,\beta /\gamma}$ and \M$_{2}$  &  $2.5_{-0.5}^{+0.7}$ \\
20 keV fixed binning for \M$_{1,\beta/ \gamma}$ and \M$_{2}$ &  $2.9_{-0.6}^{+0.8}$ \\
No crystal bulk $^{238}$U and $^{232}$Th chains             &  $2.8_{-0.5}^{+0.7}$ \\
No crystal surface $^{238}$U and $^{232}$Th chains          &  $2.8_{-0.6}^{+0.7}$ \\
No 10 mK sources $^{238}$U and $^{232}$Th chains             &  $2.2^{+0.7}_{-0.5}$ \\
MC statistics  (nuisance parameter)                         &  $2.8^{+0.7}_{-0.6}$ \\
\M$_{1,\alpha}$ range = 3000 -- 6360 keV                   &  $3.8 \pm 0.9$ \\
\noalign{\smallskip}\hline

\end{tabular}
\end{table}

We also verified that the reconstructed background index is not biased, by comparing the distribution of background indexes in toy Monte Carlo simulations to that of the reference fit. 

\subsection{Residual alpha background}
\begin{figure}[h!]
\centering
\includegraphics[width=0.45\textwidth]{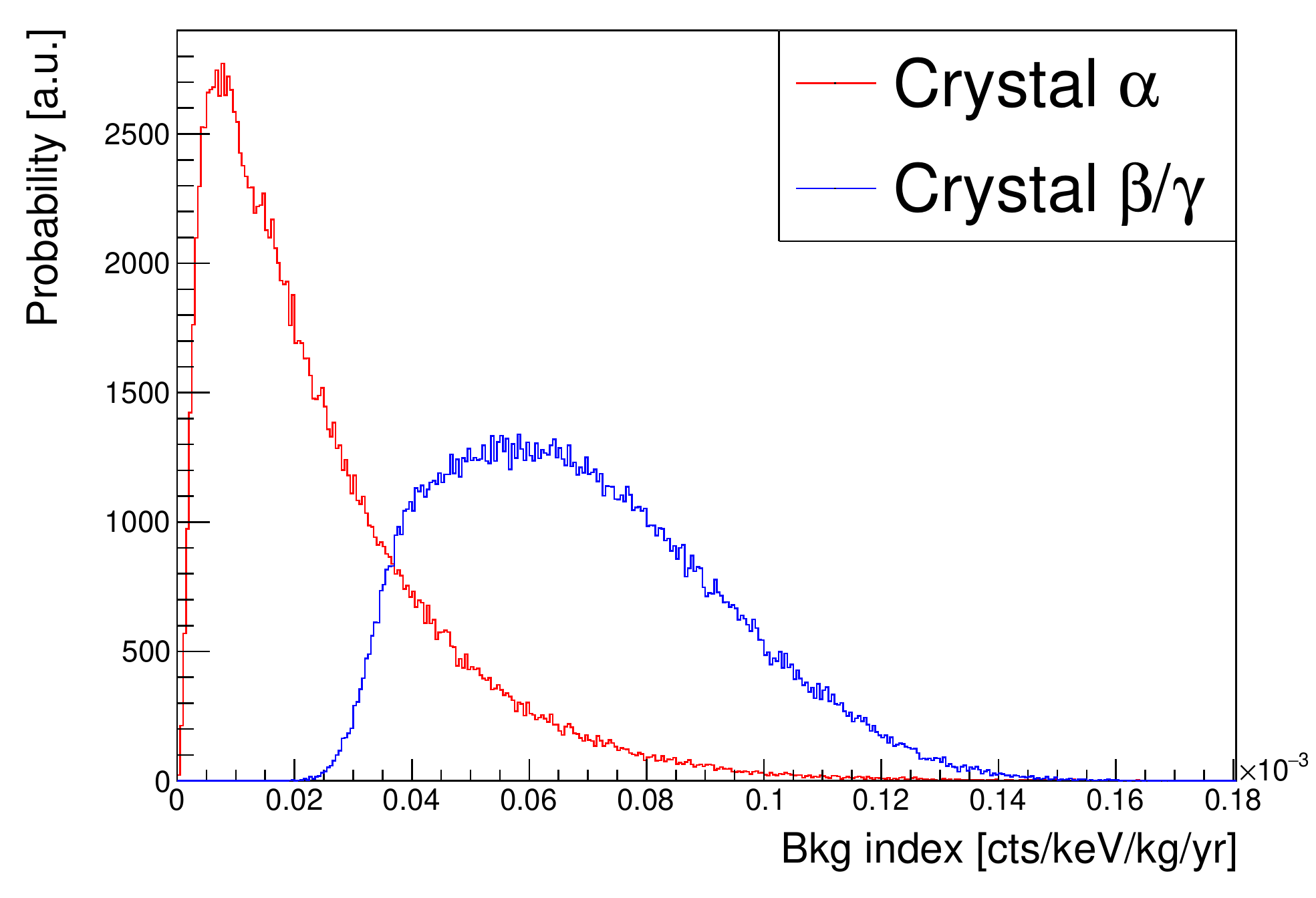}
\caption{Posterior distribution of background index of the crystal from $\alpha$ and $\beta$/$\gamma$ contamination.}
\label{alpha_beta_crystal_BI}
\end{figure}

\begin{figure*}
    \centering
    \includegraphics[width=0.45\textwidth]{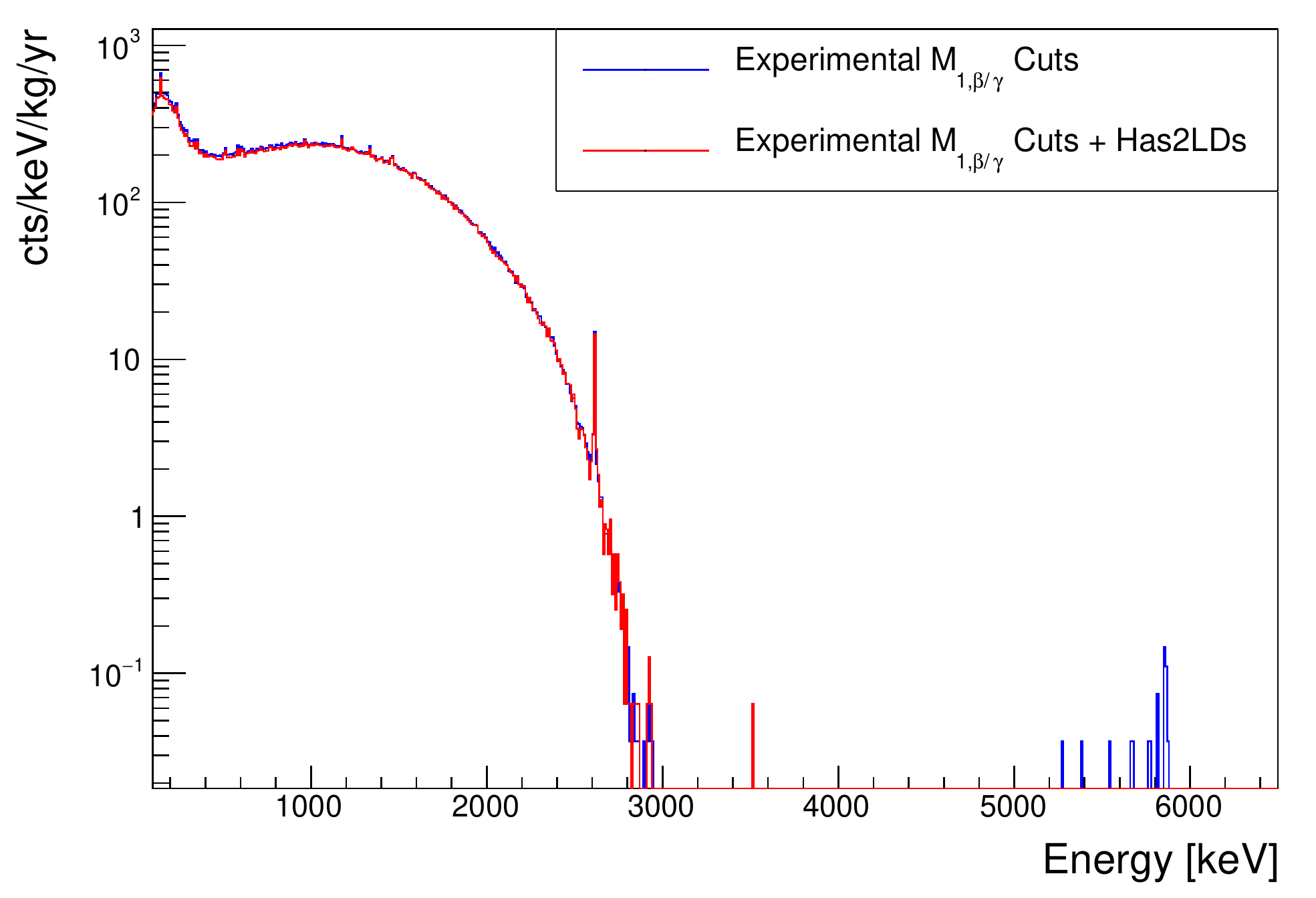} 
    \includegraphics[width=0.45\textwidth]{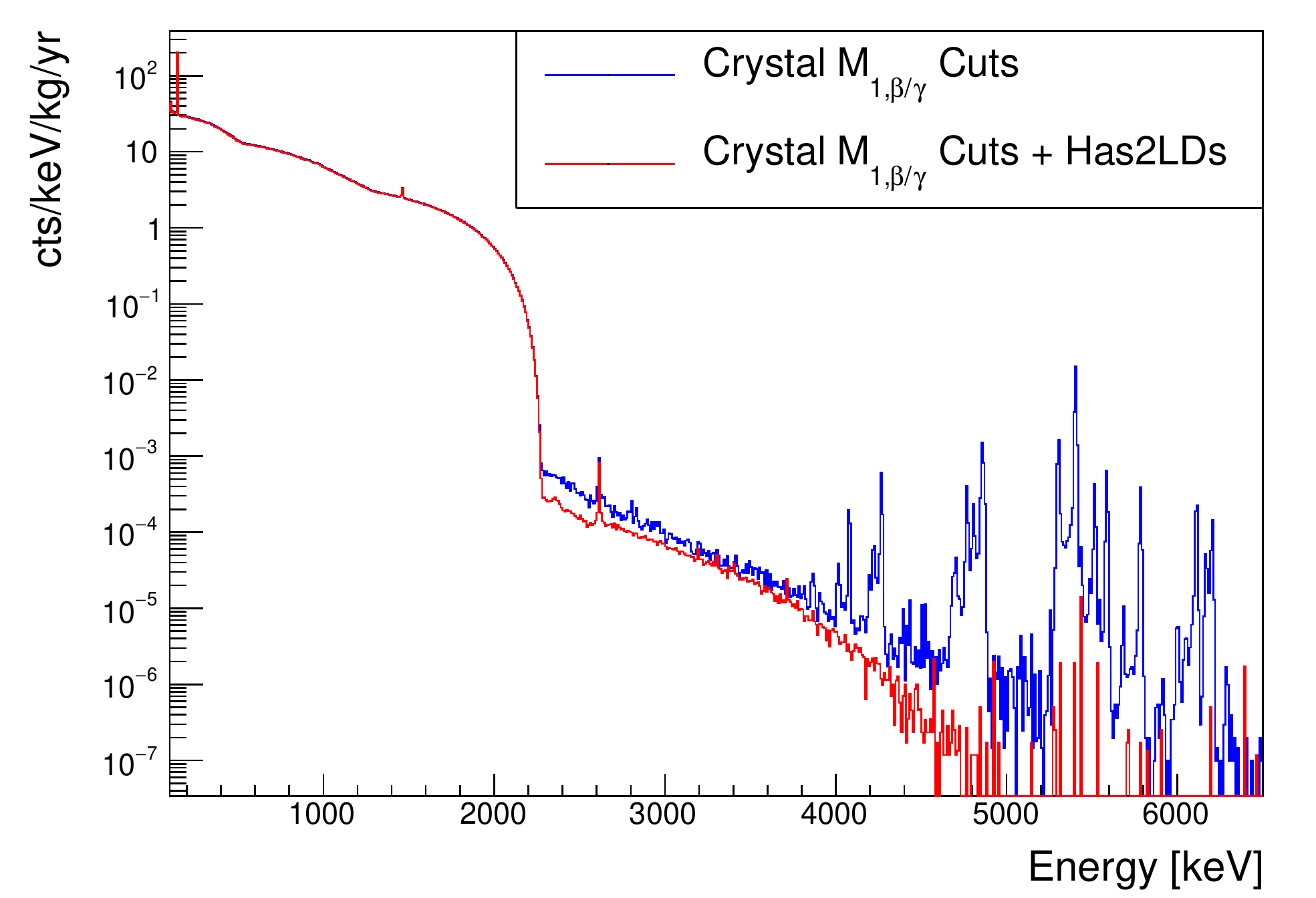}
    \caption{ {\it Left:}   Experimental \M$_{1,\beta /\gamma}$ spectrum (in blue) adding a cut to select crystals that face two LDs (in red). {\it Right:} Fit reconstruction of the crystal component from \M$_{1,\beta /\gamma}$ spectrum (in blue), adding a cut to select crystals that face two LDs (in red).} 
 \label{fig:Has2LDs}
\end{figure*}

Due to our $\alpha$ particle rejection, background events in the ROI from $^{226}$Ra and $^{228}$Th subchains in the bulk and the surface of the crystals generally arise only from energy depositions of $\beta$ or $\gamma$ particles. However, $^{238}$U, $^{234}$U, $^{230}$Th, $^{210}$Po and $^{232}$Th could also produce events in the ROI through energy deposits of $\alpha$ particles. Even if we apply a light yield cut to remove $\alpha$ background, it is still possible that some $\alpha$ events pass this selection cut.\\
 \indent From the background index distribution of the crystals, one can separate the background from $\beta/\gamma$ decays from that coming from $\alpha$ decays,  
as shown in Fig. \ref{alpha_beta_crystal_BI}.
One can observe that a non-negligible part of the crystal background index is coming from $\alpha$'s that passes the light yield cut. This $\alpha$ background is coming from surface contamination of the crystals. It corresponds to an $\alpha$ particle that deposits energy in the crystal and where the nuclear recoil deposits energy in the LD. This kind of events can pass the light yield  cut mainly for the crystals that face only one LD. We show in Fig. \ref{fig:Has2LDs}, left,  the experimental \M$_{1,\beta /\gamma}$ spectrum including all crystals, and the resulting spectrum selecting only the crystals that face two LDs. Such cut remove all the remaining alphas around 5.8 MeV. This effect is also visible in the background model. Figure \ref{fig:Has2LDs}, right, shows the reconstruction of the crystal component of \M$_{1,\beta /\gamma}$ spectrum, and the resulting spectrum selecting only the crystals that face two LDs.
We remind that in CUPID-Mo 5 of the 20 LMOs are facing only one LD due to being in the bottom floor of the towers, one LD was not operational and one had a poor performance affecting a further 4 LMOs. We expect that in the case where  all the crystals face two LDs, as in CUPID, the $\alpha$ background contribution should be negligible.

\section{Conclusion}
In this work we present the development of a background model capable of describing very accurately the CUPID-Mo experimental data with  2.71~kg $\times$ yr exposure.
We have performed a simultaneous fit of three data spectra, \M$_{1,\beta/\gamma}$, \M$_2$ and  \M$_{1,\alpha}$, to detailed Monte Carlo simulations. The model is performed in a Bayesian framework with a MCMC approach. We have shown by a fit to a $^{56}$Co calibration source that the MC implementation is accurate and that the MC is able to describe well the data. 
We used a total of 67 background sources including the bulk and surface radioactive contaminations in the crystal and  the components of the set-up. 
We have performed systematic checks varying the binning, the energy fit region and the choice of background sources that showed the stability of the model.
\\ \indent We have found that the  radiopurity of the \LMO crystals is sufficient to reach the goals of  the future \onbb experiment CUPID. The radiopurity levels of $^{226}$Ra and $^{228}$Th are below 0.5 $\mu$Bq/kg. We obtain a background index in the region of interest of  
3.7$^{+0.9}_{-0.8}$~\text{(stat)}$^{+1.5}_{-0.7}$ \text{(syst)}~$\times~10 ^{-3}$  \text{counts/$\Delta E_{\text{FWHM}}$/mol$_{\text{iso}}$/yr}, the lowest in a bolometric $0\nu\beta\beta$ decay experiment.
\\ \indent The detailing of the background achieved in this work enables promising further studies. We can obtain the \nnbb decay rate of $^{100}$Mo with high precision. It also allows for studies on various process which could distort the spectral shape, like  Bosonic neutrinos, CP violation or \onbb with Majoron(s) emission. 

\section{Acknowledgments}

This work has been performed in the framework of the CUPID-1 (ANR-21-CE31-0014) and LUMINEU programs, funded by the Agence Nationale de la Recherche (ANR, France).
We acknowledge also the support of the P2IO LabEx (ANR-10-LABX0038) in the framework ''Investissements d'Avenir'' (ANR-11-IDEX-0003-01  Project ''BSM-nu'') managed by ANR, France.

The help of the technical staff of the Laboratoire Souterrain de Modane and of the other participant laboratories is gratefully acknowledged. 
We thank the mechanical workshops of LAL (now IJCLab) for the detector holders fabrication and CEA/SPEC for their valuable contribution in the detector conception. 
F.A. Danevich, V.V. Kobychev, V.I. Tretyak and M.M. Zarytskyy were supported in part by the National Research Foundation of Ukraine Grant No. 2020.02/0011.  A.S. Barabash, S.I. Konovalov, I.M. Makarov, V.N. Shlegel and V.I. Umatov were supported by the Russian Science Foundation under grant No. 18-12-00003. J. Kotila is supported by Academy of Finland (Grant Nos. 314733, 320062, 345869).
Additionally the work is supported by the Istituto Nazionale di Fisica Nucleare (INFN) and by the EU Horizon2020 research and innovation program under the Marie Sklodowska-Curie Grant Agreement No. 754496. This work is also based on support by the US Department of Energy (DOE) Office of Science under Contract Nos. DE-AC02-05CH11231, and by the DOE Office of Science, Office of Nuclear Physics under Contract Nos. DE-FG02-08ER41551, DE-SC0011091. This research used resources of the National Energy Research Scientific Computing Center (NERSC) and the IN2P3 Computing Centre.
This work makes use of the {\it Diana} data analysis software and the background model based on JAGS,  developed by the CUORICINO, CUORE, LUCIFER, and CUPID-0 Collaborations. 
\noindent
Russian and Ukrainian scientists have given and give crucial contributions to CUPID-Mo. For this reason, the CUPID-Mo collaboration is particularly sensitive to the current situation in Ukraine. The position of the collaboration leadership on this matter, approved by majority, is expressed at \url{https://cupid-mo.mit.edu/collaboration#statement}. Majority of the work described here was completed before February 24, 2022.

\bibliographystyle{spphys}       
\bibliography{biblio}

\end{document}